\documentclass[useAMS,usenatbib]{mn2e}
\usepackage{epsfig}
\usepackage{epstopdf}
\usepackage{graphicx}
\usepackage{amssymb}
\usepackage{amsmath}
\usepackage{graphicx}
\usepackage{subcaption}
\usepackage[utf8]{inputenc}
\usepackage[export]{adjustbox}
\usepackage{wrapfig}
\usepackage{textcomp}
\usepackage[usenames,dvipsnames]{color}
\usepackage{gensymb}
\usepackage{booktabs}
\usepackage{amsmath}
\usepackage{lscape}
\usepackage[english]{babel}
\usepackage{graphicx}
\usepackage{float}
\usepackage{fixltx2e}
\title[Galactic Conformity in Cross-Correlations]{Environmental Quenching and Galactic Conformity in the Galaxy Cross-Correlation Signal}
\author[Hatfield \& Jarvis]{P. W. Hatfield$^{1}$\thanks{peter.hatfield@physics.ox.ac.uk} and M.J. Jarvis$^{1, 2}$\\
$^{1}$Astrophysics, University of Oxford, Denys Wilkinson Building, Keble Road, Oxford, OX1 3RH, UK\\
$^{2}$Department of Physics, University of the Western Cape, Bellville 7535, South Africa\\
}

\begin{document}

\date{}

\pagerange{\pageref{firstpage}--\pageref{lastpage}} \pubyear{2017}

\maketitle

\label{firstpage}

\begin{abstract}

It has long been known that environment has a large effect on star formation in galaxies. There are several known plausible mechanisms to remove the cool gas needed for star formation, such as strangulation, harassment and ram-pressure stripping.
It is unclear which process is dominant, and over what range of stellar mass. In this paper, we find evidence for suppression of the cross-correlation function between massive galaxies and less massive star-forming galaxies, giving a measure of how less likely a galaxy is to be star-forming in the vicinity of a more massive galaxy. We develop a formalism for modelling environmental quenching mechanisms within the Halo Occupation Distribution scheme. We find that at $z \sim 2$ environment is not a significant factor in determining quenching of star-forming galaxies, and that galaxies are quenched with similar probabilities when they are satellites in sub-group environments, as they are globally. However, by $z \sim 0.5$ galaxies are much less likely to be star forming when in a high density (group or low-mass cluster) environment than when not. This increased probability of being quenched does not appear to have significant radial dependence within the halo at lower redshifts, supportive of the quenching being caused by the halting of fresh inflows of pristine gas, as opposed to by tidal stripping.  Furthermore, by separating the massive sample into passive and star-forming, we see that this effect is further enhanced when the central galaxy is passive, a manifestation of galactic conformity.

\end{abstract}


\section{Introduction} \label{sec:intro}

Galaxy colours have long been observed to be bimodal, with a `red sequence' and `blue cloud', with the `green valley' separating them (e.g. \citealp{Bell2004}). Extensive work interpreting these galaxy colours has allowed a connection to be made to physical properties of the galaxies e.g. the stellar populations that make up the galaxies, their stellar mass and star formation rates (\citealp{Bruzual2003b} and many others). It is established that the red sequence corresponds to a passive `red and dead' population of massive galaxies, where star formation has ceased, and that the blue cloud corresponds to typically less massive galaxies that are still actively forming stars (\citealp{Conselice2014}) - this observational result leads naturally to the question of what causes this transition (referred to as `quenching'), and where and when does it occur for which galaxies? Temporally, global star formation is known to peak around 6-7 billion years ago, and to have dropped off since (\citealp{Madau1996,Madau1998};  see \citealp{Madau2014} for a review). However, spatially, it is also known that, beyond redshift and stellar mass, the environment a galaxy is found in can have a large effect on whether it is quenched or not, the so-called `colour-density' relation, that passive, redder galaxies are typically found in denser regions of the Universe. This holds even after controlling for the fact that more massive galaxies are biased towards denser regions, and is closely related to the `morphology-density' relation, that these passive galaxies in dense regions are typically elliptical, and the star forming galaxies in less dense regions are more likely to be spirals. This phenomenon was first investigated by \citet{Oemler1974,Davis1976,Dressler1980} and others, and now enjoys extremely robust and well studied measurements in the local Universe e.g. in the Sloan Digital Sky Survey (SDSS;\citealp{Ball2006}) and in the Two Degree Field Galaxy Redshift Survey (2dFGRS; \citealp{Balogh2004}). Extending our understanding of of the role of environment to higher redshifts ($z \sim 1-2$) is active observational programme today e.g. \citet{Darvish2017,Fossati2016}.

What processes might give rise to these relations in the galaxy population? Key processes in galaxy evolution are often classified into `nature' and `nurture' effects, e.g. internal processes such as cooling and feedback versus interactions with other galaxies and the local environment - often a variety of processes are invoked to explain environmental-based observations.  For example, many mechanisms involving removing the cool gas needed for star formation have been proposed to account for the observed quiescence in satellite galaxies around central massive galaxies. They range from strangulation (tidal effects from the gravitational potential allows the gas in the satellite to leave) and ram pressure stripping (removal of gas by `winds' in the hot intra-cluster medium) to harassment (flybys from other galaxies) - see \citet{Hirschmann2014} for more details. The mass of the dark matter halo is also believed to have a key role in these processes; \citet{Hartley2013a} suggest that because only a modest energy input is needed to keep the gas in the halo hot, there should be a critical dark matter halo mass that divides those that host passive and those that host star forming galaxies.  A study by \citet{Peng2010a} using SDSS and zCOSMOS (\citealp{Scoville2007})  take an empirical approach to modelling the build up of stellar mass in galaxies and conclude that quenching effects can be strongly distinguished into environmental quenching (e.g. satellite galaxy infall processes), merger quenching and mass quenching, an unknown process with strength proportional to the star formation rate which the authors speculate could involve active galactic nuclei (AGN) feedback. Understanding what gives rise to these independent `environmental' quenching and `mass' quenching effects is an ongoing challenge.

Although host halo mass is known to be important, it is also becoming apparent that it is not the only factor in environmental quenching. Multiple studies have reported correlations in the properties of nearby galaxies, even after accounting for halo mass, an effect now known as galactic conformity (e.g. \citealp{Hearin2015}). Galactic conformity typically manifests as correlations in the sSFR of galaxies spatially close e.g. a tendency for galaxies close together to either all be passive or all star-forming. More specifically, conformity can be broken down into 1-halo and 2-halo conformity (in analogy with the 1-halo and 2-halo terms in halo occupation distribution modelling). 1-halo conformity describes correlations within a halo, typically scales less than a megaparsec (see \citealp{Weinmann2006}) and 2-halo conformity describes correlations between nearby halos, typically on scales of more than a megaparsec, \citet{Kauffmann2013}. 1-halo conformity, although not perfectly understood, does not contradict the basic picture of dark matter halos as isolated environments; it is possible to imagine that it could be explained through galaxy interactions, AGN, and other mechanisms within the halo. 2-halo conformity is less straightforward and does not permit such a simple explanation for why correlations exist on such large scales, violating the usual assumption of the halo model that everything inside a halo only depends on the halo mass. A small 2-halo conformity would be expected from the fact that massive halos are typically clustered, and more massive halos typically contain more quenched objects, leading to objects being more passive on scales larger than their individual halos - but conformity appears to be larger than expected from this effect. \citet{Hearin2015} suggest 2-halo conformity could be explained through large-scale tidal environments affecting individual halos and/or as a consequence of assembly bias, where the assembly history of a halo has an effect on its bias and also its contents. The most simple manifestation of halo bias is the result that at fixed halo mass, halos that were assembled earlier are more highly biased. \citet{Hearin2015} suggest that the halos that are assembled earlier will contain older galaxies that are more likely to have used up their gas reservoir; these galaxies will be more highly clustered and we see this as 2-halo conformity.

Environmental quenching and the colour-density relation are typically studied either in the context of group and cluster catalogues or nearest-neighbour type density measurements. Conformity has typically been observed thus far through satellite density profiles and similar techniques e.g. \citet{Hartley2014a}. Both the colour-density relation and conformity make up part of the statistical distribution of stellar mass and star formation and thus should be detectable in the correlation function. In \citet{Hatfield2015} we used the VISTA (Visible and Infrared Survey Telescope for Astronomy) Deep Extragalactic Observations (VIDEO) survey \citep{Jarvis2013} to model the two-point correlation function and study the galaxy-halo relation. In this paper we use the same observation selection criteria and cuts, and a similar approach to the calculation of the correlation function, to investigate environmental effects. We seek to investigate this by looking at the cross-correlation signal between higher-mass and lower-mass galaxies, and then varying the star formation rates of the samples.

This paper is organised as follows: first we summarise the sample selection made in \citet{Hatfield2015}, and discuss how we measure the cross-correlation function. We then measure the cross-correlation function for a series of sub-samples split by mass, redshift and star formation rate in the survey. Finally we interpret our results in terms of environmental quenching and conformity by incorporating a quenching model into the Halo Occupation Distribution (HOD) framework.

All magnitudes are given in the AB system \citep{Oke1983} and all calculations are in the concordance cosmology $\Omega_{\Lambda}=0.7$, $\Omega_{m}=0.3$ and $H_{0}=70 \text{ km} \text{ s}^{-1} \text{Mpc}^{-1}$ unless otherwise stated.

\section{Observations} \label{sec:OBS}

In this section we describe the optical and near infrared data used to select the galaxies in our sample, and provide information on the photometric redshift, star formation rate and stellar mass estimates that underpin our analysis. The selection and data reduction is the same as in \citet{Hatfield2015} and based on the VIDEO survey described in \citet{Hatfield2015}.

\subsection{VIDEO and CFHTLS} %%%

The VIDEO Survey \citep{Jarvis2013} covers three fields in the southern hemisphere, each carefully chosen for availability of multiband data, to total 12 deg$^{2}$ when complete. The $5 \sigma$ depths of VIDEO in the four bands are $Z=25.7$, $Y=25.5$, $J=24.4$, $H=24.1$ and $K_s=23.8$ for a 2'' diameter aperture.

Our catalogue was constructed by combining the VIDEO data set with data from the T0006 release of the Canada-France-Hawaii Telescope Legacy Survey (CFHTLS) D1 tile \citep{Ilbert2006,Gwyn2012}, which provides photometry with $5 \sigma$ depths of $u^{*}= 27.4$, $g^{\prime}=27.9$, $r^{\prime}=27.6$, $i^{\prime}=27.4$ $z^{\prime}=26.1$ over 1 deg$^{2}$ of the VIDEO XMM3 tile (which will be joined by two other adjacent tiles). This data set (and the template fitting discussed in section \ref{sec:LePHARE}) has already been used in many extragalactic studies to data (e.g.  \citealp{White2015,Johnston2015,Hatfield2015}).

\subsection{LePHARE and SExtractor} \label{sec:LePHARE}

The sources in the images are identified using SExtractor, \citep{Bertin1996} source extraction software, with 2'' apertures. Photometric redshifts, stellar masses and star formation rates used in this study were calculated using LePHARE \citep{Arnouts1999,Ilbert2006}, which fits spectral energy distribution (SED) templates to the photometry. Further information on detection images used, detection thresholds and the construction of the SED templates is given in \citet{Jarvis2013}.

\subsection{Final Sample} \label{sec:final_sample}

{\sc SExtractor} identifies 481,685 sources in the 1~deg$^2$ with detections in at least one band. As outlined in \citet{Hatfield2015}, we remove sources in regions effected by excess noise and bright stars with a mask, $K_{s}>$23.5 magnitude cut, and colour cut around a stellar locus, following the approach of \cite{Baldry2010}, to remove stars. VIDEO has a $90$ percent completeness at this depth \citep{Jarvis2013} and \citet{McAlpine2012} estimate this colour cut leaves stars contributing less than 5 per cent of the sample. The final galaxy sample comprises 97,052 sources.

\begin{figure*}
\includegraphics[scale=0.3]{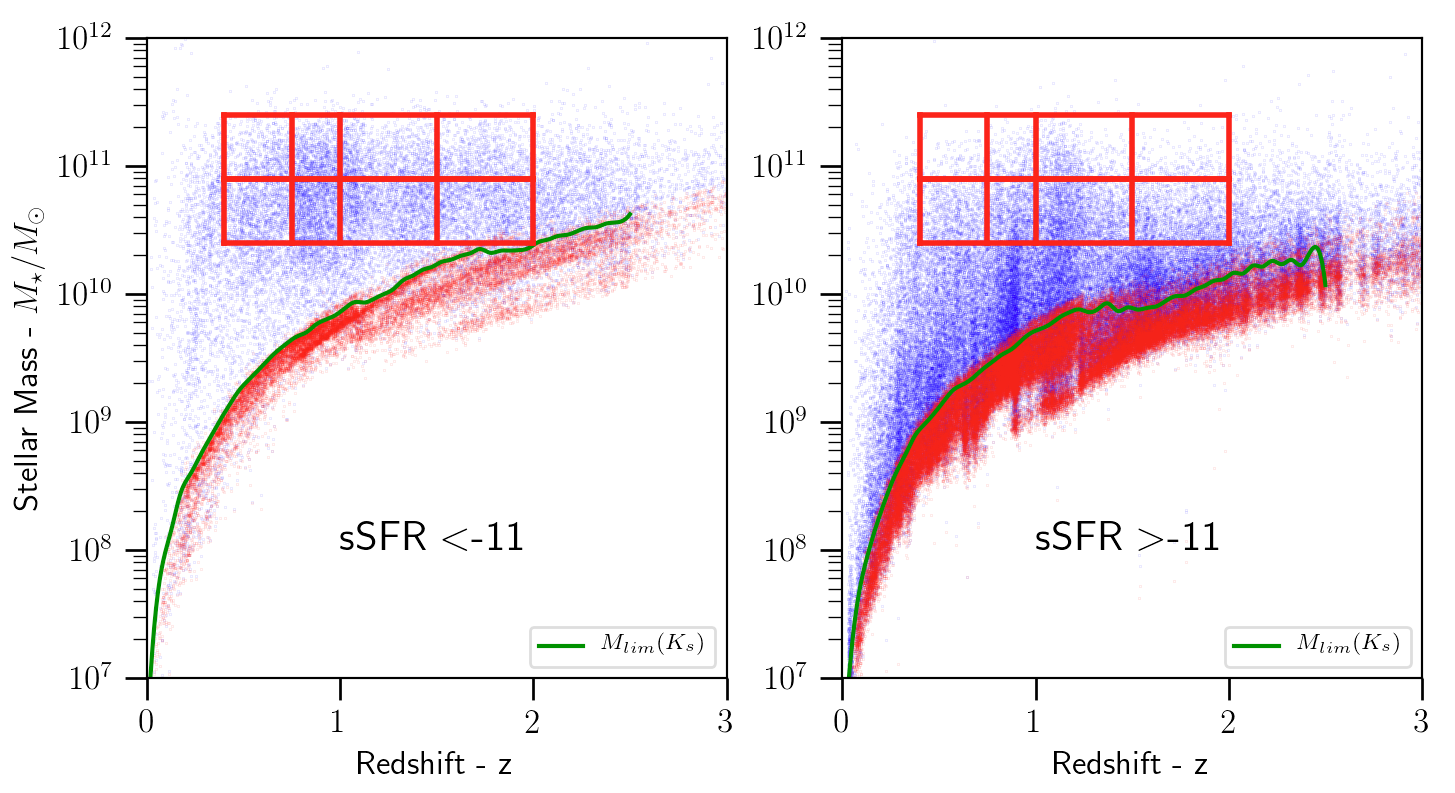}
\caption{The mass and redshift of galaxies, considered after application of the magnitude cut, star exclusion and mask, are shown here in blue. The red points mark the stellar mass limit for all objects that could be detected with our apparent magnitude limit of $K_{s}<23.5$, and the green curve the implied 90 per cent stellar mass completeness limit, following the approach of \protect\cite{Johnston2015}. The red boxes illustrate the redshift and stellar mass selected sub-samples that we consider in section \ref{sec:RESULTS},with the left plot corresponding to the passive sample, and the right hand plot the star-forming sample.}
\label{fig:mass_z}
\end{figure*}

\begin{figure*}
\includegraphics[scale=0.35]{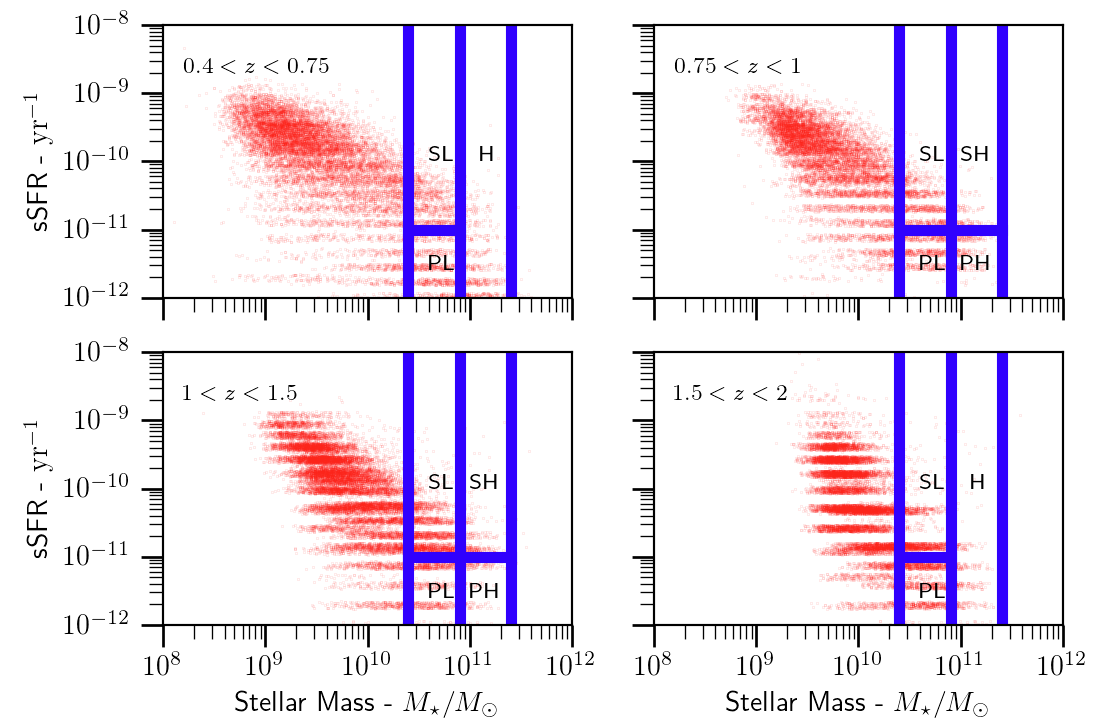}
\caption{The mass and sSFR of galaxies in our final sample, in the four redshifts considered in section \ref{sec:RESULTS}. The blue lines delineate the various samples. The initials indicate which section corresponds to each sample: `P'=passive, `S'=star forming, `H'=high mass, `L'=low mass}
\label{fig:mass_ssfr}
\end{figure*}

Figure \ref{fig:mass_z} shows our final sample in stellar mass-redshift space broken down into passive and star-forming, with the bold red lines delineating our sub-samples. Above the green curve is the part of the parameter space within which we can be very confident of completeness (see \citealp{Hatfield2015} for a discussion of the calculation of completeness limits). Conversely the red dots represent objects where incompleteness may be a factor.

Figure \ref{fig:mass_ssfr} shows a scatter plot of stellar mass versus specific star formation rate at the four redshifts considered in section \ref{sec:no_selection} and \ref{sec:ssfr_selection}, and the subsamples used - see fig.3 from \citet{Ilbert2015} for an example of similar measurements in the literature. There is likely some incompleteness at lower star-formation rates for low stellar masses in figure \ref{fig:mass_ssfr}, but not in the stellar mass and star-formation rate bins considered in this work, as the completeness curve in figure \ref{fig:mass_z} shows. When we divide by sSFR, we use $\log_{10} sSFR\lessgtr-11$, which was motivated by a combination of a need to have a large number of galaxies in each sample, as well as literature results for the divider between passive and main sequence galaxies e.g. \citet{Knobel2014} divide at $\log_{10} sSFR=-11.25$ at $M_{\star}=10^{11} M_{\sun}$. We refer to samples as being `passive' or `star-forming' based on this divider in this work, although we acknowledge that this divider could include the lower portion of the main sequence in the passive sample under some definitions. Table \ref{tab:counts} summarises the number counts and associated passive fractions. Star formation rates are derived from the templates used to construct the photometric redshifts (that is to say each template was constructed with a star formation history, and the star formation rate is that of the best fitting template). The sSFR values not taking certain values is an unphysical result of the fact that star formation rates are calculated from a discrete set of templates. See \citet{Jarvis2013} for a more in depth description of the templates used in the photometric redshift fitting.

As can be seen from Table \ref{tab:counts}, the passive fraction is $\sim 0.65$ at the lower redshifts, drops slightly in the $1<z<1.5$ redshift bin and then rises again in the $1.5<z<2$ bin. This is slightly unexpected - typically one would expect the passive fraction at fixed stellar mass to drop with redshift (e.g. \citealp{Tinker2013a}). We suggest that this high passive fraction at high redshift could be due to a) the difficulties of estimating star formation rates in dusty galaxies e.g. some of the `passive' galaxies could actually be reddened dusty star forming galaxies and b) potentially cosmic variance on the moderately small samples sizes (see Fig.7 in \citealp{Johnston2015}). When we require passive fractions in analysis in later parts of this paper, we use the more robust values (at the same stellar masses and redshifts) from \citet{Darvish2016} instead of those calculated directly from the number counts. This corresponds to marginally lower passive fraction values (but qualitatively identical redshift trends) over the first three redshift bins, as their colour cut likely corresponds to a slightly different sSFR cut, but correctly has a very low fraction at the highest redshift bin. We note that if we lower the mass limit (we are using comparatively small bins of mass-redshift parameter space) then the fraction of star-forming galaxies rapidly increases and the conventional trend is derived, in line with \citet{Darvish2016}. Number counts do not impact on the correlation function as long as the probability of being observed versus not being observed has no spatial dependence.

\begin{table*}
\caption {Number of galaxies for each redshift, stellar mass, and star formation rate bin, with the associated passive fraction for the low mass galaxies from a) the number counts and b) \citet{Darvish2016}. H denotes our high mass ($10.9<\log_{10}{(M_{min}/{\rm M}_{\odot})}<11.4$) galaxies, PL the passive low mass galaxies, and SL the star forming low mass galaxies ($10.4<\log_{10}{(M_{min}/{\rm M}_{\odot})}<10.9$, $\log_{10} sSFR\lessgtr-11$). Note that these counts are actually the sums of the weights of the galaxies in each redshift bin \citep{Hatfield2015}.} \label{tab:counts}
\begin{tabular}{@{}cccccc@{}}
\toprule
Redshift Range 	& $N_{g}$ (H)	& $N_{g}$ (PL) & $N_{g}$ (SL)	& $f_{p}$ (VIDEO) & $f_{p}$ (Darvish et al., 2016)		\\
{$0.40<z<0.75$} & 350 & 1193 & 628 & 0.655 & 0.55\\
{$0.75<z<1.00$} & 524 & 1622 & 860 & 0.653 & 0.5\\
{$1.00<z<1.50$} & 748 & 2171 & 1880 & 0.535 & 0.3\\
{$1.50<z<2.00$} & 400 & 1670 & 975 & 0.631 & 0.2\\
\bottomrule
\end{tabular}
\end{table*}

The star formation properties of galaxies in VIDEO have been explored in other works (\citealp{Zwart2014,Johnston2015}). However these works have not studied \textit{where} the star formation is occurring.

\section{The Angular Cross Correlation Function} \label{sec:selection}

\subsection{Background}

A range of statistical measurements are used to study the spatial distribution of galaxies and to characterise clustering, including nearest-neighbour, adaptive kernel smoothing and Voroni tessellation methods (reviewed in \citealp{Darvish2015}). In this study we focus on the two-point correlation function, a measure of how much more likely two galaxies are to be at a given separation than they would be for a uniform random distribution. Correlation functions are a useful probe because they are well studied, permit simple interpretations in terms of galaxy interactions and are closely related by Fourier transforms to the power spectrum. Correlation functions have already been used to probe the galaxy-halo connection in the VIDEO survey (\citealp{Lindsay2014,Hatfield2015}). 

The underlying fundamental relation is the full spatial correlation function in three dimensional space; however we only have the galaxy locations in two dimensional angular space, and limited radial location information from the photometric redshift. The angular and redshift observables can be related to the spatial correlation function either by calculating the angular correlation function, and compare to angular projections of the spatial correlation function, or to use the redshift information to form the projected correlation function in both transverse and longitudinal directions (which incorporates redshift space distortions, which are normally integrated out), see \citet{Davis1983} and \citet{Fisher1994}. Here we focus on the angular correlation function as the projected correlation function requires very precise knowledge of the redshifts of the sample to avoid being biased, and is in general more appropriate for surveys with more accurate redshifts e.g. spectroscopic surveys.

\subsection{Definition}\label{sec:ACF_def}

The angular two-point correlation function $\omega(\theta)$ is a measure of how much more likely it is to find two galaxies at a given angular separation than a uniform unclustered Poissonian distribution:

\begin{equation}
dP=\sigma(1+\omega(\theta))d\Omega ,
\end{equation}

where $dP$ is the probability of finding two galaxies at an angular separation $\theta$, $\sigma$ is the surface number density of galaxies and $d\Omega$ is solid angle. We require $\omega(\theta)>-1$ and $\lim_{\theta \to \infty}\omega(\theta) = 0$ for non-negative probabilities and for non-infinite surface densities respectively. If the two galaxies are from the same population, this is the \textit{auto} correlation function. If they are from different populations, this is the \textit{cross} correlation function.

The conventional way to estimate $\omega(\theta)$ for the auto correlation function is with the \citet{Landy1993} estimator, which is based on calculating $DD(\theta)$, the normalised number of galaxies at a given separation in the real data, $RR(\theta)$, the corresponding figure for a synthetic catalogue of random galaxies identical to the data catalogue in every way (i.e. occupying the same field) except position, and $DR(\theta)$, the number of galaxy to synthetic point pairs. \citet{Szapudi1998} generalise this to the cross correlation function with:

\begin{equation}
\omega(\theta)=\frac{D_{1}D_{2}-D_{1}R-D_{2}R+RR}{RR} ,
\end{equation}

where $D_{1}D_{2}(\theta)$ is the number of `population 1' to `population 2' pairs of galaxies at a separation $\theta$, $D_{1}R(\theta)$ is the number of `population 1' to `random' pairs etc.

The cross-correlation function can be used to demonstrate a physical association between two phenomena - or conversely the cross-correlation function can also be a useful check that two distributions are not correlated when they are not expected to be, in which case the cross-correlation should be consistent with zero. \citet{Coupon2015} for example cross-correlate galaxies at different redshifts to validate their photometric redshifts - different redshift bins should be causally separated parts of space and hence should have negligible cross-correlation. A non-zero cross correlation would correspond to cross contamination between the two bins.

The cross-correlation function has also been shown to have great potential in the context of the multi-tracer technique \citep{Seljak2009}, in which galaxy samples which trace the underlying matter distribution in different ways are used simultaneously to constrain cosmological parameters with a greatly reduced contribution to the error budget from cosmic variance. Cross-correlations can also be used to make inferences about populations with no redshift information (e.g. un-matched radio sources) when correlated with a population that does, e.g. \citet{Fine2015}.

The literature is rich with approaches to modelling the auto-correlation function. To a fairly high degree of precision, a power-law can be fitted with a gradient of $\sim-0.8$. More recently, Halo Occupation Distribution (HOD; \citealp{Cooray2002}) models of the 2-point correlation function have seen great success in galaxy evolution studies, and give physical results in agreement with other methods (e.g. \citealp{Coupon2015,Chiu2016}). Models typically prescribe the mean number of galaxies in a halo as a function of halo mass and assume a) this occupation number has a Poissonian distribution and b) that these galaxies trace some choice of dark matter halo profile (possibly with a central galaxy at the centre of the halo). Then the HOD model, choice of halo profile, halo mass function and a halo bias prescription can be translated into a spatial correlation function, and then projected to an angular correlation function. The correlation function is split into a `1-halo' term of clustering within halos, and a `2-halo' term, of clustering between halos.

On large scales, viewing galaxies as biased tracers of the dark matter distribution in the linear regime, the cross correlation function can be easily modelled. The spatial auto-correlation functions of two galaxy samples in the linear regime are $\xi_{\textrm{galAA}}=b_{\textrm{A}}^2 \times \xi_{\textrm{DM}}$ and $\xi_{\textrm{galBB}}=b_{\textrm{B}}^2 \times \xi_{\textrm{DM}}$, where $b_{\textrm{A}}$ and $b_{\textrm{B}}$ are the galaxy biases of the samples, and $ \xi_{\textrm{DM}}$ is the dark matter correlation function. The cross-correlation between the two samples on linear scales is then $\xi_{\textrm{galAB}}=b_{\textrm{A}} \times b_{\textrm{B}} \times \xi_{\textrm{DM}}$ e.g. the geometric mean. On these scales the cross-correlation function does not provide any additional information than that contained within the two auto-correlation functions e.g. $\xi_{\textrm{galAB}}^{2}=\xi_{\textrm{galAA}} \times \xi_{\textrm{galBB}}$. This can be exploited to estimate the bias of samples that are too small to measure the auto-correlation function (e.g. cross-correlate radio galaxies and IR galaxies as in \citealp{Lindsay2014}, cross-correlate galaxies and galaxy-groups as in \citealp{Knobel2012}).

\citet{Simon2008} show how to extend the HOD approach to the cross-correlation function with the joint halo occupation distribution (JHOD) model. Within conventional HOD, the 1-halo term is constructed by finding the contribution to the clustering from each halo mass, and weighting by the halo mass function (HMF). The contribution from each halo mass is the convolution of the halo profile with itself, weighted by $\langle N^2 \rangle$, where $N$ is the average number of galaxies in a halo of that mass. For a cross-correlation function, this weighting is instead replaced with $\langle N_A N_B  \rangle$, where $N_A$ and  $N_B$ are the average numbers of galaxies of type A and B in a halo of that mass. Effectively the two auto-correlation functions provide information about $\langle N_A^2 \rangle$ and $\langle N_B^2 \rangle$, and the cross correlation gives you information about $\langle N_A N_B \rangle$ i.e. on small scales the cross-correlation \textit{does} provide information not contained within the two auto-correlation functions ($\xi_{\textrm{galAB}}^{2} \neq \xi_{\textrm{galAA}} \times \xi_{\textrm{galBB}}$). In particular it gives you the \textit{covariance} of the two occupations. This can be scaled to lie in [-1,1] e.g. for a given halo mass, members of the two samples are never found together (-1), occupy the halos independently (0), or are always found together (+1). Its impact is on the 1-halo term of the cross-correlation function, to enhance it when galaxies are always found in the same halo, and dampen it when the galaxies are rarely found in the same halo.

\citet{Simon2008} fitted the auto and cross correlation functions for red and blue galaxies simultaneously. They also investigated the effect of allowing red and blue galaxies to lie on different profiles, in which case the 1-halo cross-correlation function is proportional to the convolution of these different halo profiles. These two approaches (`JHOD' introducing covariance into how the galaxies occupy the halos, and altering the profiles the different samples followed) have the same prosaic effect of modelling different galaxy types having an affinity (or not) to be nearby, but are mathematically distinct; \citet{Simon2008} conclude that the two models were broadly degenerate for the levels of precision explored in their work.

\subsection{Application to Environmental Quenching and Conformity} \label{sec:environment_tracer}

As discussed in section \ref{sec:intro}, the star formation rate of galaxies is believed to be tied to the environment within which they live e.g. \citet{Woo2012} and \citet{Gabor2014} track how the quenched fraction changes as a function of radial location in the halo and the halo mass. This, as part of the description of the spatial distribution of galaxies, must play a role in determining the correlation function, but is not typically included in modelling of correlation functions. Most HOD studies typically just make measurements as a function of luminosity/stellar mass e.g. \citet{Coupon2015,Hatfield2015}. Some (e.g. \citealp{Coupon2012,McCracken2015}) model the auto-correlation of the whole galaxy population, and then the passive galaxy population, finding that the passive galaxies are typically found in higher mass galaxies, but is not typically linked to measurements of the passive fraction etc. Related to the idea of environmental quenching, conformity refers to the observation that nearby galaxies have correlated star formation rates. Again, this is an observation about spatial distributions of galaxies, so correlation functions seem well suited to describe them.

We suggest a potentially valuable way of observing environmental effects (including conformity) in what we call here the `environment tracer method'. We take a `tracer' population, which in some sense is believed to trace, or even cause, some particular environment, and a `follower' population, which has its properties predominantly decided by the environment it finds itself in. In this paper we choose the tracer population to be massive galaxies, as only these have the ability to have a large impact on their environment, and lower mass galaxies as the follower population. We then measure the cross-correlation between the tracer and follower populations as a function of their properties e.g. the specific star formation rate of each population. This allows us to see if the properties of the follower population are effected by the properties of the tracer population - comparatively lower power in the cross-correlation means the follower population is inhibited by the tracer population, comparatively high power in the cross-correlation means the follower population is preferentially found near the tracer population. We note that the cross-correlation function is formally symmetric between the two populations; the distinction between tracer and follower population is purely one made for ease of discussion and comparison.

We can investigate for example how other measurements of conformity translate into the environment tracer method. In figure 10 of \citet{Weinmann2006}, the fraction of satellite galaxies in a halo found to be late or early is measured as a function of halo mass, for a variety of samples.  They found that, at fixed halo mass, the late (early) fraction increased when the central itself was late (early). The natural equivalent to look at this through correlation functions is to see the impact on the cross correlation of massive galaxies and less massive galaxies, when the massive sample is split by type (early/late). We would expect a reduction in the amplitude of the cross correlation from massive lates to low mass early types, and vice versa. In figure 2 of \citet{Kauffmann2013}, the median sSFR of galaxies as a function of a distance from a massive central galaxy is measured. They find that the median sSFR increases up to radii of 4~Mpc when the sSFR of the central is increased. The environmental tracer method equivalent again is to make the central massive galaxies the tracer population and the other galaxies the follower population. We would then expect the cross-correlation function to be enhanced when the populations have correlated star formation rates.

\section{Results} \label{sec:RESULTS}

Our approach is to measure galaxy interactions/the effect galaxies can have on each other through the environment tracer method. When we calculate $RR$ for our estimation of the correlation function we use 500,000 random data points in the 1 square degree. For the uncertainties on the correlation functions, as per \citet{Hatfield2015}, we use 100 bootstrap resamplings to estimate the uncertainty by taking the 16th and 84th percentiles of the resampling. We also, as in \citet{Hatfield2015}, estimate the correlation function a) using the kernel smoothing method discussed in that paper b) using the weights system of \citet{Arnouts2002} to account for uncertainty in the redshift estimates (shown by \citealp{Asorey2016a} to reduce bias) and c) treating the integral constraint as part of the model.

\subsection{The High- to Low-mass Cross-Correlation Function} \label{sec:no_selection}

In this section we use the environment tracer method as described in section \ref{sec:environment_tracer}, with higher mass galaxies ($10.9<\log_{10}{(M_{min}/{\rm M}_{\odot})}<11.4$) as the tracer population, and lower mass galaxies ($10.4<\log_{10}{(M_{min}/{\rm M}_{\odot})}<10.9$) as the follower population. We then measure the cross correlation function between the tracer and follower populations, and sub-divide the follower population by star formation rate ($\log_{10} sSFR\lessgtr-11$). We do this for four redshift bins: $0.4<z<0.75$, $0.75<z<1.00$, $1.00<z<1.50$ and $1.50<z<2.00$. Redshift range choices were motivated by the desire to obtain a similar number of sources in each bin.

\begin{figure*}
\includegraphics[scale=0.6]{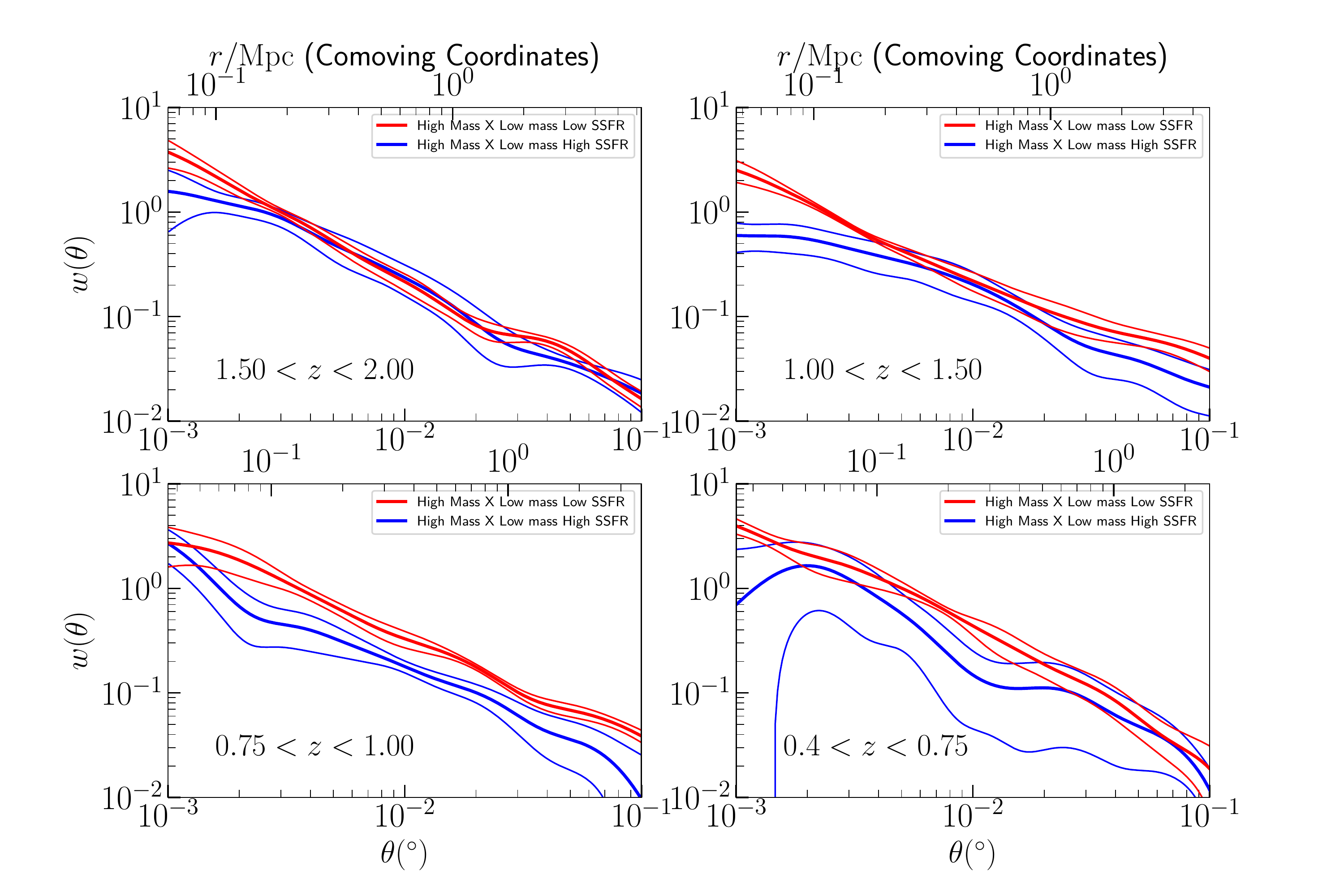}
\caption{The cross-correlation function signal between low mass ($10.4<\log_{10}{(M_{min}/M_{\odot})}<10.9$) and high mass ($10.9<\log_{10}{(M_{min}/M_{\odot})}<11.4$) galaxies in VIDEO in four redshift bins. The low mass sample is selected to be passive ($\log_{10} sSFR<-11$) for the red curves, and star forming ($\log_{10} sSFR> -11$) for the blue curves. At the highest redshift, the two curves are very similar; the role of environment in determining star formation rate seems to be small. At $1< z<1.5$ the passive low-mass galaxies are more clustered around the massive galaxies, but only on small scales. In the $0.75<z<1$ and $0.4<z<0.75$ bins, the enhancement has reached the larger scales, and is substantial on all scales.}
\label{fig:quenching}
\end{figure*}

In figure \ref{fig:quenching} each subplot shows the cross-correlations for a different redshift bin. Each curve is the cross correlation between our `tracer' massive galaxy sample, and a low mass `follower' sample. The red curve shows the cross-correlation when the low-mass sample is passive, and the blue curve star forming. 

In our highest redshift bin ($1.5 < z < 2$), although the error bars are large, there is no significant difference between the two curves, on all but the smallest scales. This indicates that, although there may be different number densities of the passive and star forming low mass samples, there is no significant difference in their geometric distribution around massive galaxies, both within the halo of the massive galaxy, and in the nearby halos i.e. there appears to be no environmental effect on star formation in the low mass galaxies, apart from the central $\sim 0.1$Mpc around the massive galaxy.\footnote{In section \ref{sec:final_sample} we note that the passive fraction derived from the number counts was unexpectedly high for the high redshift bin. We believe that this is predominantly a result of using small mass-redshift parameter space bins, and that our SFR estimates remain largely accurate up to $z\sim3$ (see \citealp{Johnston2015}). Nonetheless, if, as mentioned in section \ref{sec:final_sample}, dust is a non-trivial unaccounted for factor at high redshift, then we can only make the weaker statement of red and blue galaxies being equally clustered, as opposed to passive and star-forming galaxies, for the $1.5<z<2$ redshift bin. }. This result is robust to moderate changes ($\sim 20$ percent) in choice of sSFR cut, so choice of cut does not unduly influence our conclusions.

At the intermediate redshifts, $1 < z < 1.5$, there is substantial enhancement of the cross-correlation with the passive low-mass sample, compared with the star forming low-mass sample, out to a projected radius of ~0.2 Mpc, roughly the length scale of the halos hosting the massive galaxy sample. We interpret this as a clear signal of quenching associated with the environment of the massive galaxy sample.

However for the lowest two redshift bins, we see the enhancement extends to all scales that we can probe with the 1deg$^2$ field (albeit with large error bars in our lowest redshift range). By this stage environmental quenching has had such an effect that, although the average host halo mass was originally the same for passive and low-mass star-forming galaxies, they have been preferentially quenched in higher mass haloes, leaving the average halo mass higher for the passive tracer sample, imparting the higher bias evident in the higher two-halo term.

Note we have chosen to attempt to measure the cross-correlation at different redshifts for the same galaxy properties e.g. we are looking at galaxies of the same stellar mass and sSFRs at $1.5 < z < 2$ as $0.4 < z < 0.75$. This means we are not tracing the `same system' over time, as a galaxy of $M_{\star} \sim 10^{11} {\rm M}_{\sun}$  at $z \sim 2$ will have a substantially larger mass at $z \sim 0.5$. The logical alternative to our approach would be to use evolution models motivated by simulations and theory to attempt to track systems over time. We chose not to take this approach mainly on the grounds that the underlying physical process behind quenching to which we want to probe should be redshift independent e.g. statements like ``a galaxy of star formation X has its star formation reduced by a factor of Y when it enters environment Z" should not depend on the time at which the event occurs. A given process may be more or less prevalent at different redshifts, but underlying physics of the processes should not change.

\subsection{Selection by SFR in the Massive Sample} \label{sec:ssfr_selection}

We now split our environment tracer sample by specific star formation rate. This is more challenging, as the star-forming fraction of the massive sample is much smaller, using the same divider of $\log_{10} sSFR\lessgtr-11$.  We measure the effect just in our two middle redshift bins ($0.75<z<1.00$ and $1.00<z<1.50$) to explore this effect, due to number count constraints.

\begin{figure*}
\includegraphics[scale=0.6]{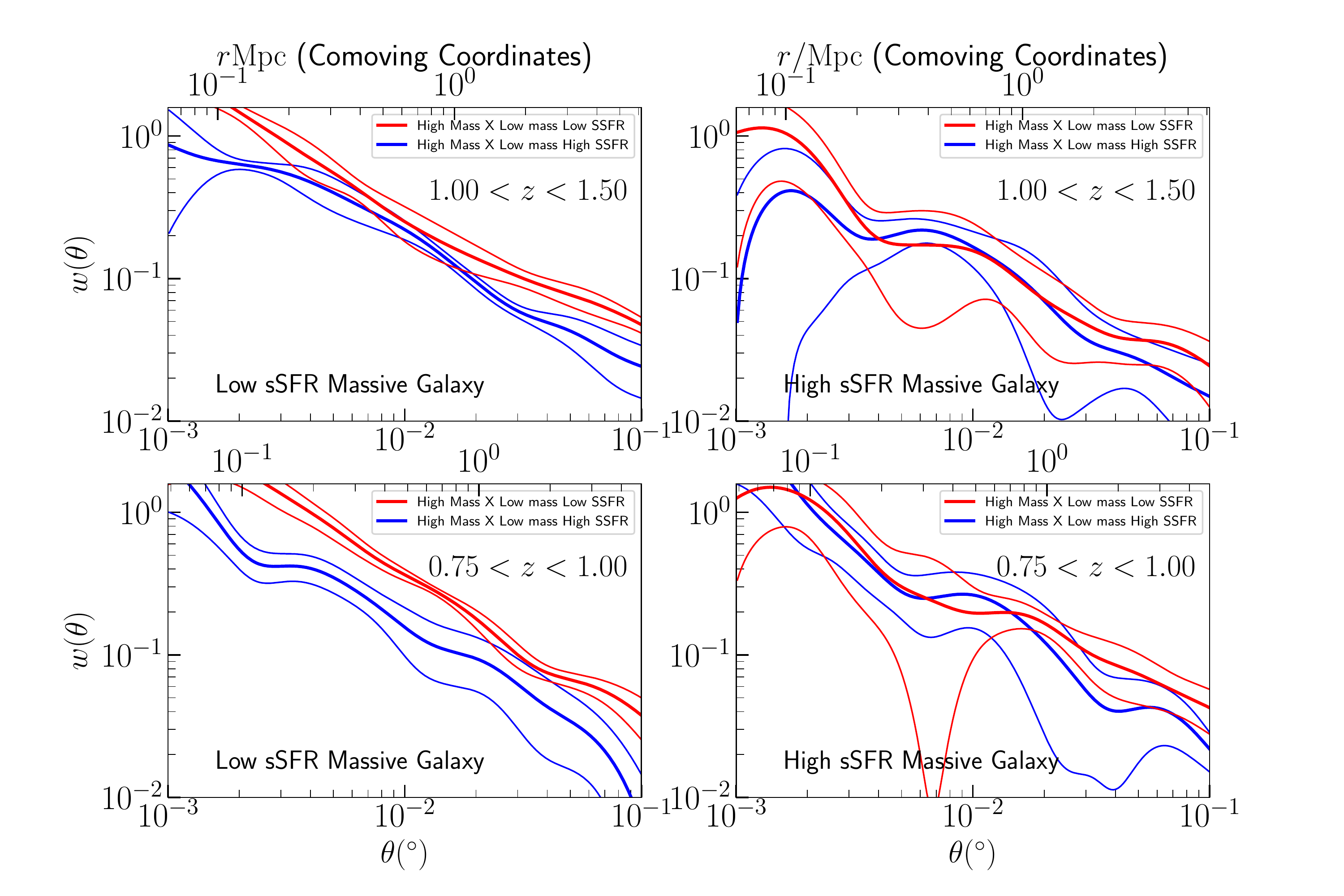}
\caption{The cross-correlation function signal between low mass ($10.4<\log_{10}{(M_{min}/M_{\odot})}<10.90$) and high mass ($10.9<\log_{10}{(M_{min}/M_{\odot})}<11.4$) galaxies in VIDEO at two redshifts. Red denotes when the low mass galaxy is passive ($\log_{10} sSFR<-11$), blue denotes starforming ($\log_{10} sSFR>-11$). In the left two plots the massive galaxy is passive, and the right two the massive galaxy is starforming (with criterion $\log_{10} sSFR\lessgtr-11$).}
\label{fig:conformity}
\end{figure*}

In fig. \ref{fig:conformity}, the top two panels are at $1.00<z<1.50$, and the bottom two sub-plots at $0.75<z<1.00$. Each subplot is the same as in Fig. \ref{fig:quenching} (e.g. the red corresponds to passive low mass sample and blue corresponds to the star forming, low mass sample), except there is also sSFR dependency in the massive sample. In the left two panels the massive sample is selected to have a low sSFR, and in the right two panels, is selected to have high sSFR.

At both redshifts, when the massive central galaxy is passive, we see qualitatively similar behaviour to that presented in fig \ref{fig:quenching} - the passive massive samples have passive low-mass galaxies more clustered around them than star-forming low-mass galaxies. However there is essentially no enhancement when the massive galaxy is star-forming - lower-mass galaxies are equally clustered around them, suggesting that they have a similar passive fraction when near a massive star forming galaxy to the total population. In other words, at fixed central stellar mass, the lower mass satellites have lower star formation rates if the central is passive, to if the central is star forming. Furthermore, it appears that the size of the effect increases from $1.00<z<1.50$ to $0.75<z<1.00$.

This illustrates what conformity looks like in cross-correlation functions. Massive passive galaxies have lower mass passive galaxies much more clustered around them than lower mass star forming galaxies. Massive star-forming galaxies appear to have no star formation rate dependence for how satellites are clustered around them. In Section~\ref{sec:model_directly} we describe a way to advance our understanding of the role of the environment, while also controlling for halo mass, by incorporating these effects into the HOD formalism.

Note that this \textit{is} a measurement of 2-halo conformity at $z\sim 1$, in the sense that there are correlations between star formation rates of galaxies on scales larger than individual halos. However it is \textit{not} (yet) evidence that this is above and beyond just what one would expect from the fact that massive halos are typically near to other massive halos and galaxies are typically more passive in massive halos. Further modelling is needed to investigate if there is a conformity effect suggestive of large scale quenched areas from assembly bias or tidal environmental effects on intermediate scales between individual haloes and the larger linear scales.

\subsection{Modelling Quenching Effects Directly} \label{sec:model_directly}

Given these results, we seek a way of modelling them within the HOD framework. We note that the two innovations of \citet{Simon2008}, allowing covariance of the red and blue galaxy occupation numbers, and allowing red and blue galaxies to trace different profiles, are both ways of modelling red and blue galaxies `not liking' being together; effectively an interaction term where galaxies are more likely to be passive when nearby to other passive galaxies - in other words conformity.  Our approach is to model the impact of this interaction on the correlation function from a quenching mechanism directly. We briefly discuss a toy model building on \citet{Simon2008} here.

\subsubsection{Radial Dependence of Quenching Within a Halo}  \label{sec:model_develop}

\begin{figure*}
\includegraphics[scale=0.4]{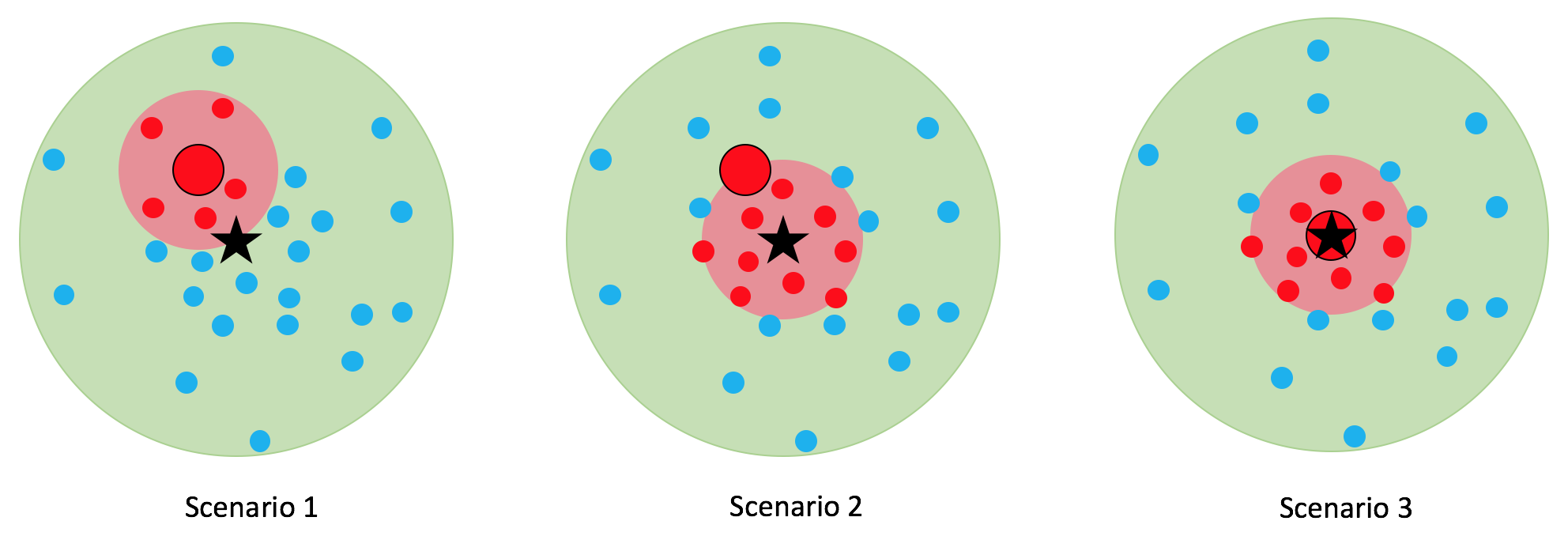}
\caption{Illustration of the three possible ways for quenching to have spatial dependence within a halo. In each figure, the pale green symbolises the extent of the dark matter halo. The black star represents the centre of the halo. The large red circle represents a high mass galaxy, that may or may not be affecting the star formation rate of lower mass galaxies, denoted by the smaller dots. The shaded pale red area denotes `quenched areas'. The lower mass galaxies are passive (denoted by being red) when in the quenched area, and star forming (denoted by being blue) when not. In Scenario 1, galaxies are preferentially quenched when near the massive galaxy, and the massive galaxy traces the halo profile. In Scenario 2, quenching is associated with proximity to the centre of the halo, and again all galaxies trace the dark matter profile. In Scenario 3, the massive galaxy is located at the centre of the halo, and quenching is associated with proximity to the centre of the halo, which coincides with the massive galaxy.}
\label{fig:diagram}
\end{figure*}

A reasonable fiducial model for the location dependence of the probability of a galaxy being quenched is that it either depends on radial distance from the centre of the halo, or it depends on distance to some other quenching source (a more massive galaxy, an AGN etc.). Three distinct scenarios (illustrated in figure \ref{fig:diagram}) then arise in the context of our earlier example:

\begin{itemize}
  \item Quenching depends on the separation of both galaxies, and both populations trace the halo profile
  \item Quenching depends on the radial distance to the centre of the halo, both populations trace the halo profile
  \item Massive galaxy/AGN sits at centre of the halo, impossible to distinguish between quenching coming from the massive object, or from radial distance to the centre of the halo
\end{itemize}

We denote the probability of being quenched in either of these scenarios as $Q(r)$, and use the parametrisation:

\begin{equation}
Q(r)=  K+\frac{S-K}{2}  \left(1- \textrm{erf}\left(\frac{\log_{10}(r)-log(r_{Q})}{a}\right) \right) ,
\end{equation}

with $0<K<1$, $0<S<1$, $0<r_{Q}$, $0<a$. This parametrisation is chosen to have galaxies quenched with probability $S$ when they are at the source of the quenching, and quenched with probability $K$ when they are far away. The length scale $r_{Q}$ describes the scale of this transition, and $a$ how sharp this transition is. Correspondingly, the probability of being star forming is $1-Q(r)$.

In conventional HOD modelling, galaxies are assumed to trace the profile of the halo, which we take to be a standard Navarro-Frenk-White (NFW, \citealp{Navarro1996}) profile.

The 1-halo term is then proportional to the convolution of the profiles each sample traces. If a galaxy is a `central' then its `profile' is taken to be a Dirac delta at the centre of the halo (this is the distinction between the central-satellite and satellite-satellite parts of the 1-halo term). We now discuss how to model each of our quenching scenarios within this framework:

Scenario 1 is the only scenario which is not equivalent to just modifying the profiles the samples trace - the other two are still symmetrical around the centre of the halo. The convolution of two NFW profiles is the probability of finding two galaxies at that separation. This pairing is then a massive-passive pair with probability $Q(r)$ and massive-SF with probability $1-Q(r)$. Thus we have for the cross correlation of the massive environmental tracer population to low-mass passive galaxies:

\begin{equation}
1+\xi_{1h}(\textbf{r}) \propto Q(\textbf{r}) \int_{\mathbb{R}^3} \rho_{\textrm{DM}}(\textbf{s}) \rho_{\textrm{DM}}(\textbf{r}-\textbf{s}) d\textbf{s} ,
\label{eq:quenching_formula}
\end{equation}

and for the cross correlation of massive to low mass star forming galaxies:

\begin{equation}
1+\xi_{1h}(\textbf{r}) \propto (1-Q(\textbf{r}) )\int_{\mathbb{R}^3} \rho_{\textrm{DM}}(\textbf{s}) \rho_{\textrm{DM}}(\textbf{r}-\textbf{s}) d\textbf{s} .
\label{eq:quenching_formula}
\end{equation}

Calculations involving NFW profiles such as these are commonly done in Fourier space, as an analytic expression for a NFW profile exists (\citealp{Scoccimarro2001}), and the convolution can be done as a multiplication. Note that this has a knock-on effect to the profiles the actual galaxies trace, with galaxies towards the centre of the halo preferentially being quenched (assuming that $Q$ increases for small radii!). The passive low mass galaxies will then trace a profile

\begin{equation}
\rho_{\textrm{passive}}(r) \propto \rho_{\textrm{DM}}(r) \int_{\mathbb{R}^3} Q(\textbf{s} )\rho_{\textrm{DM}}(\textbf{r}-\textbf{s}) d\textbf{s}
\label{eq:quenching_formula}
\end{equation}

and the star-forming low mass galaxies a profile

\begin{equation}
\rho_{\textrm{SF}}(r) \propto \rho_{\textrm{DM}}(r) \int_{\mathbb{R}^3} (1-Q(\textbf{s}))\rho_{\textrm{DM}}(\textbf{r}-\textbf{s}) d\textbf{s} .
\label{eq:quenching_formula}
\end{equation}

This is due to the fact that the probability of the low mass galaxy being at a given point in the halo is proportional to the NFW profile. Then, once it is `placed' there, the probability of being quenched is proportional to the integral over all space of the probability of being quenched, given all the possible places the massive galaxy can be.

Note that these expressions describe the global profiles of galaxies in halos of the given mass - individual halos will not necessarily be symmetric. Because of this complication, the auto-correlation is \textit{not} just the convolution of this profile. To see why, imagine one dimensional halos in a one dimensional universe. Half the halos have two galaxies, each at a radius of 1 megaparsec away from the centre of the halo (e.g. at -1 and 1 if 0 is the centre of the halo). The other half of the halos have some sort of physical process that makes galaxies be at a radius of 0.5 megaparsecs (at -0.5 and 0.5). The true one-halo auto-correlation function of this universe would have peaks at 1 and 2 megaparsecs, the only physical separations galaxies are found at. However the `profile' has galaxies at -1,-0.5, 0.5 and 1. Taking the convolution of this would leave one to the mistaken conclusion that the auto-correlation function should have peaks at 0.5, 1, 1.5 and 2. This is another way of saying that averaging to find a profile and convolution do not commute. The correct way to find the auto correlation function in the 1D model is to find the auto correlation function for each halo, and then average, as opposed to averaging the halos, and then finding the auto correlation function.

Scenario 2 and Scenario 3 can be modelled by altering the profiles the lower mass galaxies lie on. Now they are just probabilistically passive or star forming depending on their radial location in the halo e.g. $\rho_{\textrm{passive}}(r) \propto Q(r) \rho_{\textrm{DM}}(r)$ and $\rho_{\textrm{SF}}(r) \propto (1-Q(r)) \rho_{\textrm{DM}}(r)$. Then when the massive galaxy also traces the halo (Scenario 2), we obtain:

\begin{equation}
1+\xi_{1h}(\textbf{r}) \propto  \int_{\mathbb{R}^3} Q(\textbf{s}) \rho_{\textrm{DM}}(\textbf{s}) \rho_{\textrm{DM}}(\textbf{r}-\textbf{s}) d\textbf{s} ,
\label{eq:quenching_formula}
\end{equation}

for the correlation with the passive low-mass population, and

\begin{equation}
1+\xi_{1h}(\textbf{r}) \propto  \int_{\mathbb{R}^3} (1-Q(\textbf{s})) \rho_{\textrm{DM}}(\textbf{s}) \rho_{\textrm{DM}}(\textbf{r}-\textbf{s}) d\textbf{s} ,
\label{eq:quenching_formula}
\end{equation}

for the for the correlation with the star forming low-mass population. When the massive galaxy is always found at the centre of the halo (Scenario 3), the second NFW profile in the convolution collapses into a Dirac delta, and the terms become:

\begin{equation}
1+\xi_{1h}(\textbf{r}) \propto Q(\textbf{r}) \rho_{\textrm{DM}}(\textbf{r}) \label{eq:quenching_formula}
\end{equation}

for the massive central to passive low-mass galaxies cross-correlation and

\begin{equation}
1+\xi_{1h}(\textbf{r}) \propto  (1-Q(\textbf{r})) \rho_{\textrm{DM}}(\textbf{r}) .
\label{eq:quenching_formula}
\end{equation}

for the massive central to star forming low-mass galaxies cross-correlation. All three scenarios of course come with several assumptions. They all assume that the two low mass samples lie on `complementary' distributions e.g. the sum of their profiles adds to an NFW. Scenario 1 assumes that there is only one galaxy causing quenching within the halo, in practice there would be higher-order corrections for the small fraction of halos containing a higher number of quenching galaxies.

Note that in all three cases, the quenching mechanism gives a natural prediction for the fraction $f$ of quenched low mass galaxies:

\begin{equation}
f= \frac{\int r^2\rho_{\textrm{passive}}(r) \textrm{d}r}{\int r^2\rho_{\textrm{passive}}(r) \textrm{d}r + \int r^2\rho_{\textrm{SF}}(r) \textrm{d}r} .
\label{eq:pass_activ_frac}
\end{equation}

This gives a fairly natural way to `unify' the two models described in \citet{Simon2008}, as the modified profiles dictate the expected number of galaxies in the halo when another galaxy sample is present. For example, the case where the occupation numbers are different, but the passive and active galaxies still trace the same profile, is described by $S=K$, and the relative amplitude between the two profiles changes, without changing the shape of their profile.

\begin{table*} 
\begin{tabular}{ c c c c } 
 ---- & Scenario 1 & Scenario 2 & Scenario 3 \\ 
Massive Profile & $\rho_{\textrm{DM}}$ & $\rho_{\textrm{DM}}$ & $\delta$ \\  
Passive Low Mass Profile & $\rho_{\textrm{DM}} \times Q\ast \rho_{\textrm{DM}}$ & $Q \rho_{\textrm{DM}}$ & $Q \rho_{\textrm{DM}}$   \\
Star Forming Low Mass Profile & $\rho_{\textrm{DM}} \times (1-Q) \ast \rho_{\textrm{DM}}$ & $(1-Q) \rho_{\textrm{DM}}$ & $(1-Q) \rho_{\textrm{DM}}$   \\
$1+\xi_{1hP}$ (Cross) & $Q \times \rho_{\textrm{DM}} \ast \rho_{\textrm{DM}}$ &  $ (Q \rho_{\textrm{DM}}) \ast \rho_{\textrm{DM}}$ & $Q \rho_{\textrm{DM}}$ \\
$1+\xi_{1hSF}$ (Cross) & $(1-Q )\times \rho_{\textrm{DM}} \ast \rho_{\textrm{DM}}$ & $((1-Q) \rho_{\textrm{DM}}) \ast \rho_{\textrm{DM}}$ & $(1-Q) \rho_{\textrm{DM}}$ \\
$1+\xi_{1hP}$ (Auto) & NA &  $(Q \rho_{\textrm{DM}}) \ast (Q \rho_{\textrm{DM}})$ & $(Q \rho_{\textrm{DM}}) \ast (Q \rho_{\textrm{DM}})$ \\
$1+\xi_{1hSF}$ (Auto) & NA & $((1-Q) \rho_{\textrm{DM}}) \ast ((1-Q) \rho_{\textrm{DM}})$ & $((1-Q) \rho_{\textrm{DM}}) \ast ((1-Q) \rho_{\textrm{DM}})$
\end{tabular}

\caption{Summary of the resulting profiles, along with auto- and cross-correlation functions for different quenching mechanisms within a halo. $\delta$ denotes a Dirac delta, $\ast$ denotes a 3D convolution}
\label{table:convolutions}
\end{table*}

\begin{figure*}
\includegraphics[scale=1.2]{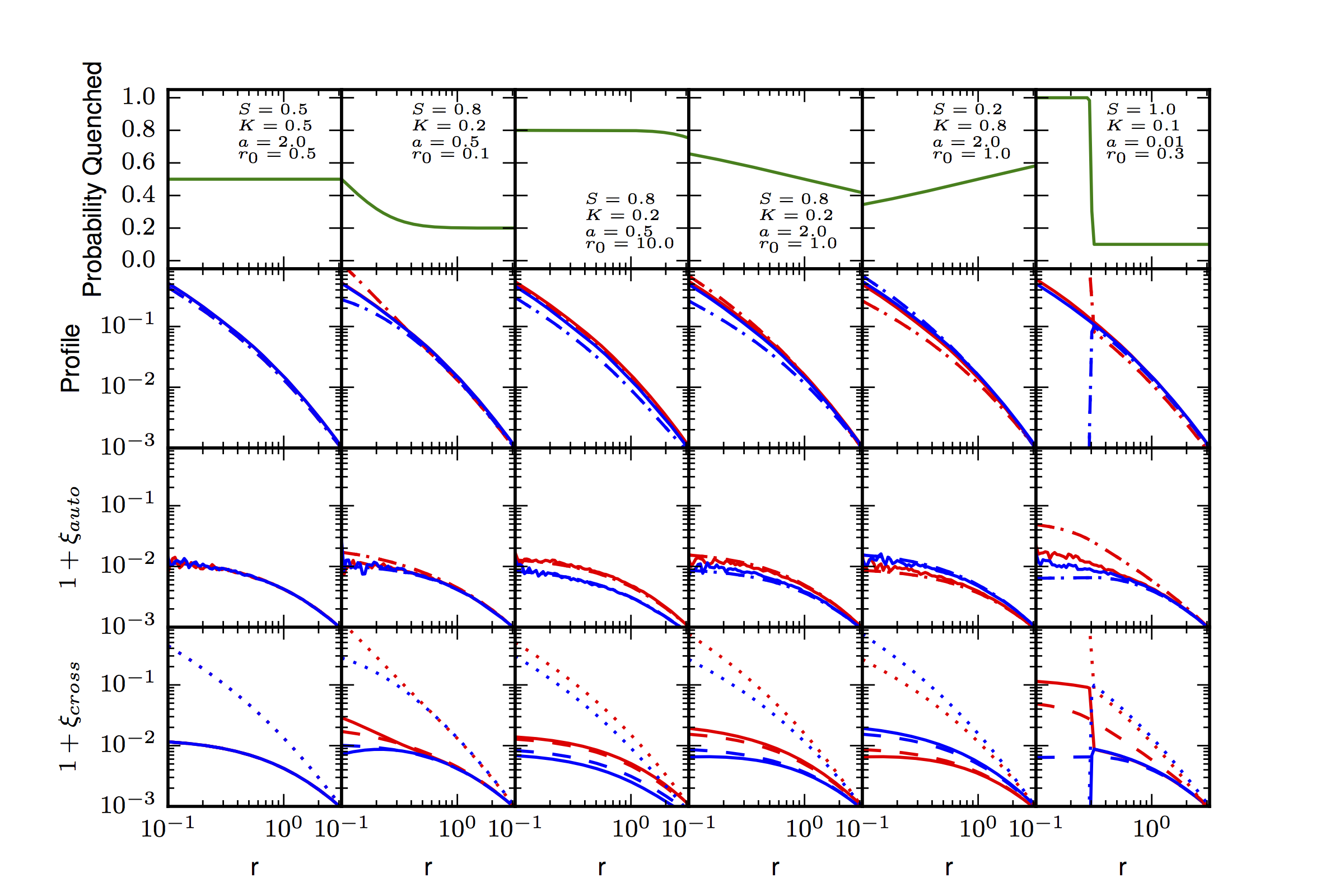}
\caption{An illustration of the consequences of different quenching models within our formalism. First row: probability of being quenched as a function of spatial separation. Second row: profile of galaxies within a halo. Third row: auto correlation of galaxy samples. Fourth row: cross correlation of low mass to high mass galaxy samples. Full lines correspond to Scenario 1, dashed lines to Scenario 2, dotted lines to Scenario 3. All functions (described in table \ref{table:convolutions}) normalised, red corresponds to passive and blue to star forming where appropriate. The differences between the Scenarios are clearly much more prominent in the cross-correlations than the auto-correlations for all models.}
\label{fig:conformity_illustration}
\end{figure*}

We summarise the profiles, auto-correlations and cross-correlations for each Scenario in table \ref{table:convolutions}. Figure \ref{fig:conformity_illustration} shows the three scenarios for several models. We use the same halo model for each plot, scale radius 1Mpc, and compactness parameter 10.

Note that necessarily if a profile or correlation function is larger on small scales it must be smaller on large scales, as they are normalised to have total weight of unity. All the panels in Fig.~\ref{fig:conformity_illustration} are normalised, and the cross-over sometimes occur out of the range plotted. However typically the amount by which the flatter distribution is greater than the more compact one on large scales is tiny, as it adds up to a greater probability due to it being weighted by $r^{2}$ when integrated over the whole volume.

The auto-correlation functions for Scenario 1 as discussed above do not present a straightforwards analytic expression that the authors can identify. However, it is still possible to calculate, by a Monte-Carlo method. We sample a location within the halo for the massive galaxy to be. We then sample two locations for the lower-mass galaxies to be, and sample whether they are passive or star forming. When a pair of the correct type is found, their spatial separation is recorded, which essentially allows DD to be constructed. The auto-correlations for Scenario 1 in Figure \ref{fig:conformity_illustration} are constructed in this manner.

In the first quenching model in Figure \ref{fig:conformity_illustration}, there is no radial dependence for quenching. The profiles the galaxies trace are NFW, the auto-correlation function is its self-convolution, and the cross correlations are the NFW self-convolution in Scenario 1 and 2, and the convolution of NFW and a point for Scenario 3 (the standard central-satellite term.

In the second quenching model, there is a slight increased tendency for galaxies to be quenched at small separations (either from each other, or from the centre of the halo depending on the scenario). It can clearly been seen that on small scales passive galaxies become more concentrated at the centre of the halo. This has corresponding effects on the auto and cross-correlations, both being enhanced/depressed on small scales. The third quenching model shows the same, but the characteristic length scale for quenching is a lot larger, and the enhancement can be seen to hold on larger scales.

The fourth quenching model shows a perhaps more realistic model, giving measurables closer to the data. In particular, we can see that, Scenario 1 has a larger impact on the differences between the cross-correlations than on the auto correlations, and Scenario 2 has a larger impact on the auto correlation differences.This shows potentially how they could be distinguished. The fifth quenching model is identical to the fourth, except swapped, in the sense that galaxy star formation is enhanced when close to other galaxies/the centre of the halo, and, as expected, we just get the inverted result.

The sixth model shows an extreme case, where galaxies are quenched with certainty at separations closer than the characteristic quenching scale. This causes the profile of the star forming galaxies to go to zero below a certain radius in Scenario 2 and 3, as no galaxies can be star forming in the centre. The auto correlations are enhanced/depressed to different degrees, as normal. But the cross correlations show the most interesting case. For Scenario 1, the cross correlation for star forming galaxies goes to zero, as massive galaxies are never seen near to star forming galaxies. Similarly for Scenario 3, but in Scenario 2, there is still some cross-correlation, as it is still possible to find massive galaxies near star forming galaxies when they are both away from the centre of the halo.

This illustrates that, within the context of this model, judicious measurements of auto and cross correlation functions can allow one to distinguish between different quenching powers, models and scenarios.

What is the physical interpretation of the three scenarios? In Scenario 1 quenching is dependent on physical proximity between the galaxies. Detection of quenching within a halo being geometrically distributed in this manner could reasonably be interpreted as supportive of quenching mechanisms like splashback, tidal stripping, harrassment, an AGN heating up its vicinity, and so on. Conversely, for Scenario 2, quenching is dependent on location within halo, and the correlation functions being supportive of Scenario 2, could be interpreted as quenching mechanisms like ambient temperature in the halo heating the gas in the galaxy. Scenario 3 is the case when the two coincide geometrically because the massive galaxy is at the centre of the halo, which makes it harder to distinguish mechanisms in the simple version of the model discussed here.

The reason in this work we do not also fit for the autocorrelations is that there will be contributions from other halo masses, which complicates matters because a) the quenching mechanism is likely halo and stellar mass dependent b) the number of quenching sources is no longer fixed at one. However we propose that the future direction of inquiry attempts to model the cross and auto correlation functions simultaneously, which we discuss further in section \ref{sec:future}.

The language used in this paper is motivated by results such as \citet{Woo2012}, who suggest that quenching is associated with location in a halo, or proximity to another galaxy. However we note that it may be possible to get similar geometric arrangements of galaxies without there being a physical interaction causing this effect - perhaps passive satellite galaxies were in more massive halos before they were accreted, and that when these halos become sub halos they are more likely to be concentrated towards the centre of the larger halo by some mechanism of the sub-halo dynamics. Furthermore, our ``characteristic quenching scale" is purely a `measured' dependence, which is not necessarily the same as the physics of the actual mechanism because the motion of the satellites within the halo could `smear' the signal e.g. suppose an AGN quenches star formation in every galaxy within 0.1~Mpc, e.g. \citet{Rawlings2004}, galaxies will be observed to be quenched at larger radii than that because their orbit might take them close to the AGN, permitting them to be quenched, but then move them further away again.

Note that this formalism still strictly describes quenching and conformity within a halo. \citet{Hearin2015} suggest their decorated HOD model may be able to explain 2-halo conformity though. In addition, our quenching parameters are strictly dependent on the sSFR cut to divide into star forming and passive, and will in general be different for different sSFR and stellar mass cuts. In a scenario where galaxies were passive at the centre of a halo, and got progressively more star forming towards the outside, one might expect $r_{Q}$ to increase as one increased the sSFR cut.

\subsubsection{The Halo Model}

Halos are non-linear small scale overdensities in the dark matter density field caused by collapse when an overdensity reaches a critical amount - in this sub-section we discuss the model and parametrisation of halo profiles that we use in our model. Both theoretical considerations and N-body simulations suggest that NFW profiles (\citealp{Navarro1996}) describe halo density profiles well. NFW profiles are parametrised by a scale radius $R_{s}$ which denotes scale of transition from the inner $r^{-1}$ density profile and the outer $r^{-3}$ density profile, a virial radius within which the dark matter has virialised, a compactness parameter which denotes the radius between the two, and a characteristic density which links to the halo mass. As the untruncated NFW profile has infinite total mass (is unnormalisable), halo models typically assume that galaxies only lie within the virial radius and that the halo density drops to zero at this point. Note that it is necessary to make some sort of similar assumption, otherwise the probability of finding galaxies within some radius is proportional to the mass enclosed within this radius, which is finite compared with the infinite mass as the radius extends to infinity. The total mass of the halo is then the density integrated over the sphere with radius $R_{vir}$.

There are multiple different prescriptions describing how halos of different masses and compactnesses link to overdensities in the global dark matter field \citep[e.g. see][for an overview]{Coe2010}. We use the parametrisation described in \citet{Coupon2012}. In this formalism, the critical overdensity $\Delta(z)$ required for virialisation is taken from \citet{Weinberg2003}, the relationship between the scale radius and the virial radius is given by $R_{vir}=c \times R_{s}$, with $c$ a function of halo mass and redshift as given in \citet{Coupon2012}.

\subsubsection{Fitting Quenching Models} \label{sec:MCMC}

We seek to model the environmental role of quenching to our observations using our results for the 1-halo term we developed in section \ref{sec:model_develop}. The theory as developed there only describes the 1-halo term resulting from galaxy interactions for a single halo mass. To fully model the cross-correlation function this must be generalised (as in conventional HOD) to include all halo masses, which is likely not straightforward as the quenching power is likely to be a function of halo mass, and the mass of the galaxy it is acting on. However our massive galaxies live in massive halos, where the HMF is dropping off quickly, which allows us to approximate that the 1-halo term is dominated by galaxies in a small range of halo masses. 

Models based solely on Scenario 1 and 2 produced extremely poor fits and were in general unable to reproduce our results, suggesting the massive galaxy must typically be at the centre of the halo (one can see by eye in figure \ref{fig:conformity_illustration} that if there is no central galaxy the one-halo term is very shallow at small radii). This is consistent with the results of \citet{Hatfield2015} that $\sim 90$ per cent of galaxies of these stellar masses at these redshifts are central galaxies. The general expression for the one-halo central-satellite term is

\begin{equation}
1+\xi(r)=\int_{M_{vir}}^{\infty} n(M) \frac{\langle N_{cen} \rangle \langle N_{sat} \rangle}{n_{1}n_{2}}  \rho(r,M) \textrm{dM} 
\label{eq:quenching_formula_reference}
\end{equation}

(see \citealp{Coupon2015}, where we remove the factor of 2, as we are considering cross correlations and don't need to account for each pair being counted twice), where $n(M)$ is the normalised HMF in counts per cubic comoving megaparsec per unit mass, $\rho$ is the normalised density of the profile per cubic comoving megaparsec, $n_{i}$ is the comoving number density of each type of galaxy, $\langle N_{cen} \rangle$ is the expected number of centrals, $\langle N_{sat} \rangle$ the expected number of satellites.

Because the satellite number rises as a power law, and the HMF drops off as an exponential, the integral is dominated by a relatively small range of halo masses, so we can take the profile dependence on halo mass outside the integral, and not consider the complication of halo mass dependence on quenching power etc. This leaves:

\begin{equation}
1+\xi(r) \approx \int_{M_{vir}}^{\infty} n(M) \frac{\langle N_{cen} \rangle \langle N_{sat} \rangle}{n_{1}n_{2}}  \textrm{dM}  \times \rho(r,M_{dom}) , 
\label{eq:quenching_formula_2}
\end{equation}

where $M_{dom}$ is the dominant halo mass probed. We take $\rho(r)$ to be a normalised NFW, modified as per Scenario 3, and calculate the integral assuming the halo occupation measured in \citet{Hatfield2015}, modifying the $n_{i}$ appropriately. In particular we take $\langle N_{sat} \rangle = f \times N_{HOD}$, where f is the implied fraction of satellite galaxies quenched according to equation \ref{eq:pass_activ_frac} and $N_{HOD}$ is the number of satellites in a halo of that mass implied by the HOD fits in \citet{Hatfield2015}. If we were doing a simultaneous fit of both occupation numbers \textit{and} a quenching mechanism, we could take $n_{i}$ and $N_{HOD}$ to be properties of the model, but this ansatz will suffice for our purposes.

Equation \ref{eq:quenching_formula_reference} shows amplitudes of the profiles depend on the passive fraction of the population as a whole, and the passive fraction of the galaxies in the vicinity of the massive galaxy. In general for our measurements, the overall amplitude is set by the halo mass the interactions are taking place in, and the ratio $R$ between the two cross-correlations in the 1-halo term is set by:

\begin{equation}
R=\frac{ f^{P}_{E} / f^{P}_{G} }{ f^{SF}_{E} / f^{SF}_{G} } =  \frac{ f^{P}_{E} / f^{P}_{G} }{ (1- f^{P}_{E}) / (1- f^{P}_{G}) }  , 
\label{eq:quenching_formula_3}
\end{equation}

where  $f^{P}_{E}$ is the passive fraction of the low mass galaxies in the environment traced by the massive galaxies, $f^{P}_{G}$ the same fraction globally (e.g. for all low mass galaxies), and $f^{SF}_{E}$ and $f^{SF}_{G}$ the corresponding star forming fractions.

Once our one-halo cross-correlation term is constructed, we then add the two-halo term. We then project into angular space using the source redshift distribution and subtract the integral constraint, described in \citet{Beutler2011}, i.e.. the integral constraint is part of the model.

In total we have seven parameters to describe each individual set of measurements: two large scale cross biases, four parameters to describe the galaxy interactions, one parameter for halo mass. We use emcee\footnote{http://dan.iel.fm/emcee/current/} (\citealp{Foreman-Mackey2012}) to provide a Markov chain Monte Carlo sampling of the parameter space to fit our correlation functions. We use a uniform prior over $0.5<b_{\times 1}^{2}<9$, $0.5<b_{\times 2}^{2}<9$, $11<\log_{10}{(M_{\mathrm{h}}/M_{\odot})}<15$, $-2<\log_{10}{r_{Q}}<0$, $0<S<1$, $0<K<1$ and $0<a<3$ (that is to say uniform in log space for mass and squared space for the biases). We used 20 walkers with 1000 steps, which have starting positions drawn uniformly from the prior.

Our likelihood is calculated using $\chi^{2}$ from both clustering measurements:

\begin{equation}  \label{eq:chi}
\begin{split}
\chi^{2}= \sum\limits_{i} \frac{[\omega_{\mathrm{pas}}^{\mathrm{obs}}(\theta_{i})-\omega_{\mathrm{pas}}^{\mathrm{model}}(\theta_{i})]^{2}}{\sigma_{w_{i_{\mathrm{pas}}}}^{2}} +  \sum\limits_{i} \frac{[\omega_{\mathrm{SF}}^{\mathrm{obs}}(\theta_{i})-\omega_{\mathrm{SF}}^{\mathrm{model}}(\theta_{i})]^{2}}{\sigma_{w_{i_{\mathrm{SF}}}}^{2}} , 
\end{split}
\end{equation}

where $\omega_{\mathrm{pas}}^{\mathrm{obs}}$ is the observed massive to passive cross correlation function data, $\omega_{\mathrm{pas}}^{\mathrm{model}}$ is the corresponding model correlation function, $\sigma_{w_{i_{\mathrm{pas}}}}$ is the error on these measurements, $\theta_{i}$ are values of theta fitted for, separated by more than the smoothing length used to estimate the correlation function (with SF corresponding to star forming for the other variables).

\subsubsection{Results of the quenching model} \label{sec:quenching_model_results}

\begin{figure*}
\includegraphics[scale=0.4]{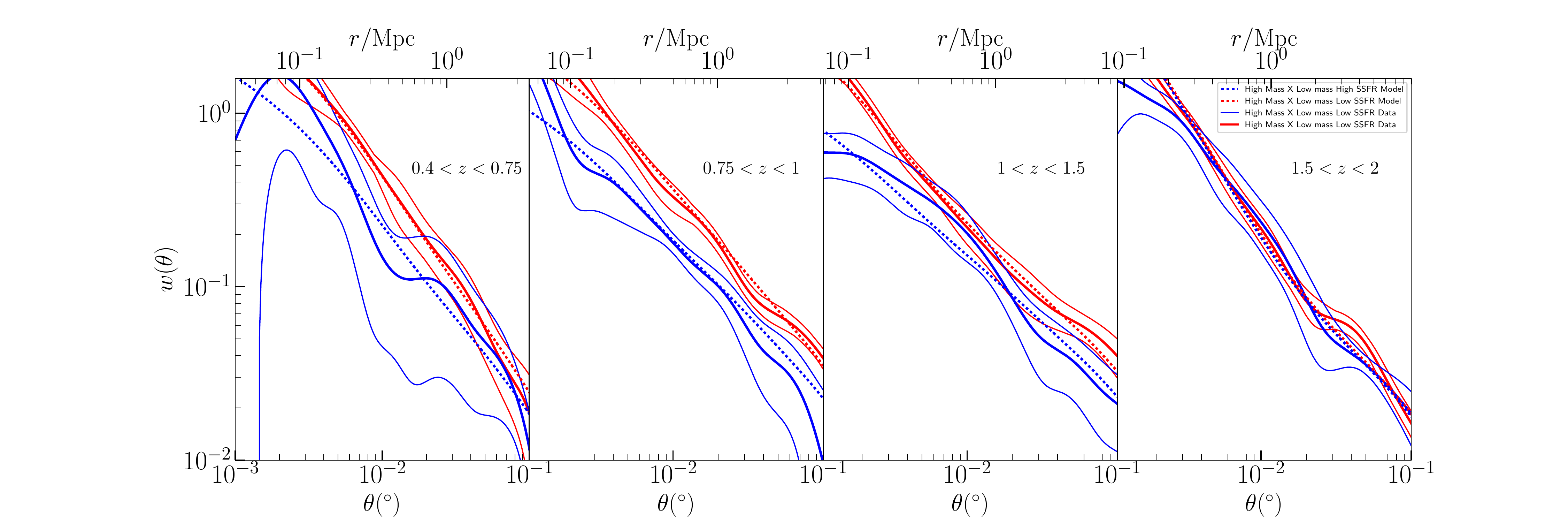}

\caption{Our `environment tracer' measurements and best fit model at our four redshift ranges. The solid lines are the data; the cross correlations of massive to less massive galaxies, red when the low mass galaxy is passive and blue when star forming. The error bars are the 16th and 84th percentiles of the estimator. The dashed lines correspond to the best fit model using the quenching modification to the 1-halo term. }

\label{fig:fit_example}
\end{figure*}

\begin{figure*}
\includegraphics[scale=0.4]{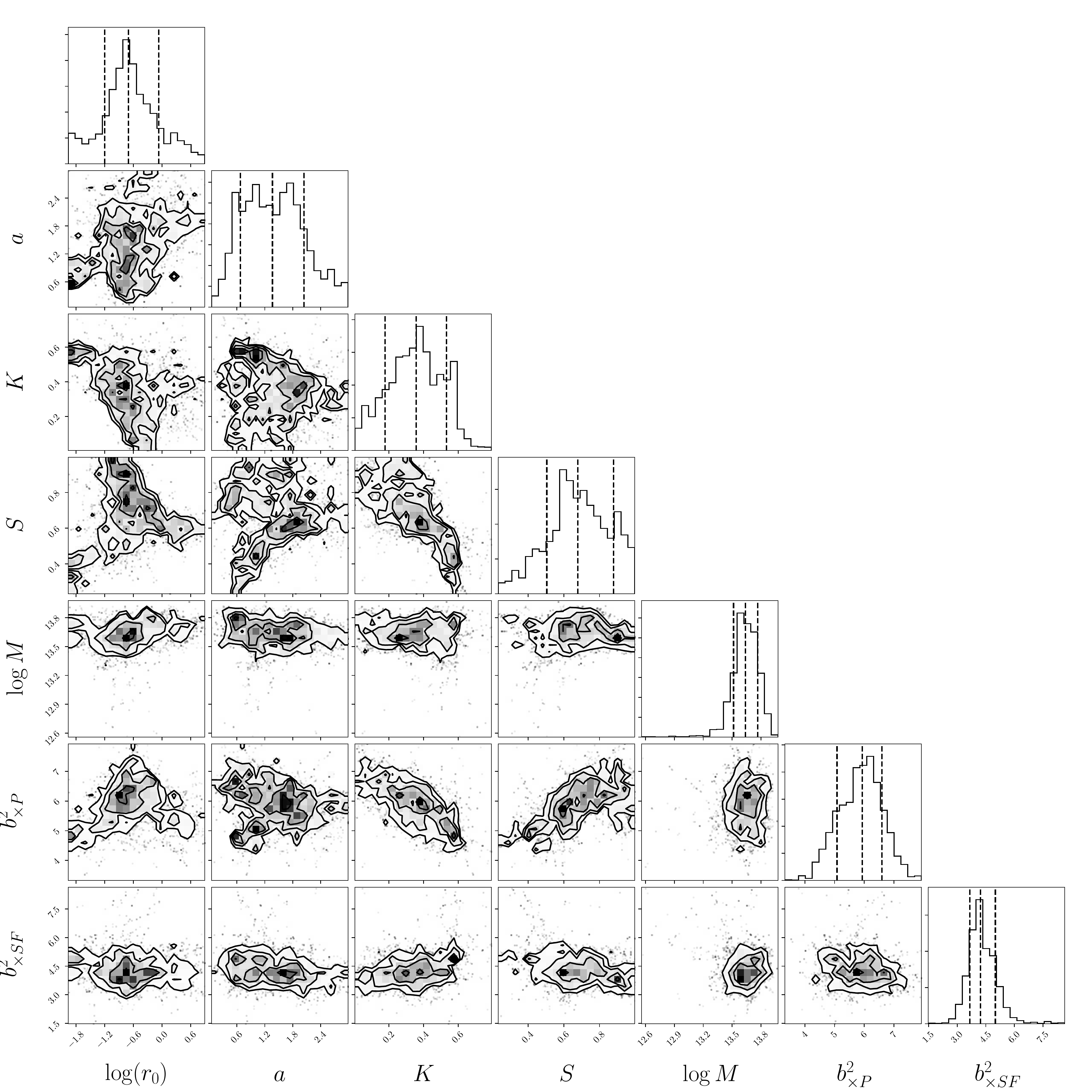}
\caption{Corner plot of our posterior for the parameters of our quenching model at $z=1.24$}
\label{fig:triangle}
\end{figure*}

\begin{figure*}
\includegraphics[scale=0.4]{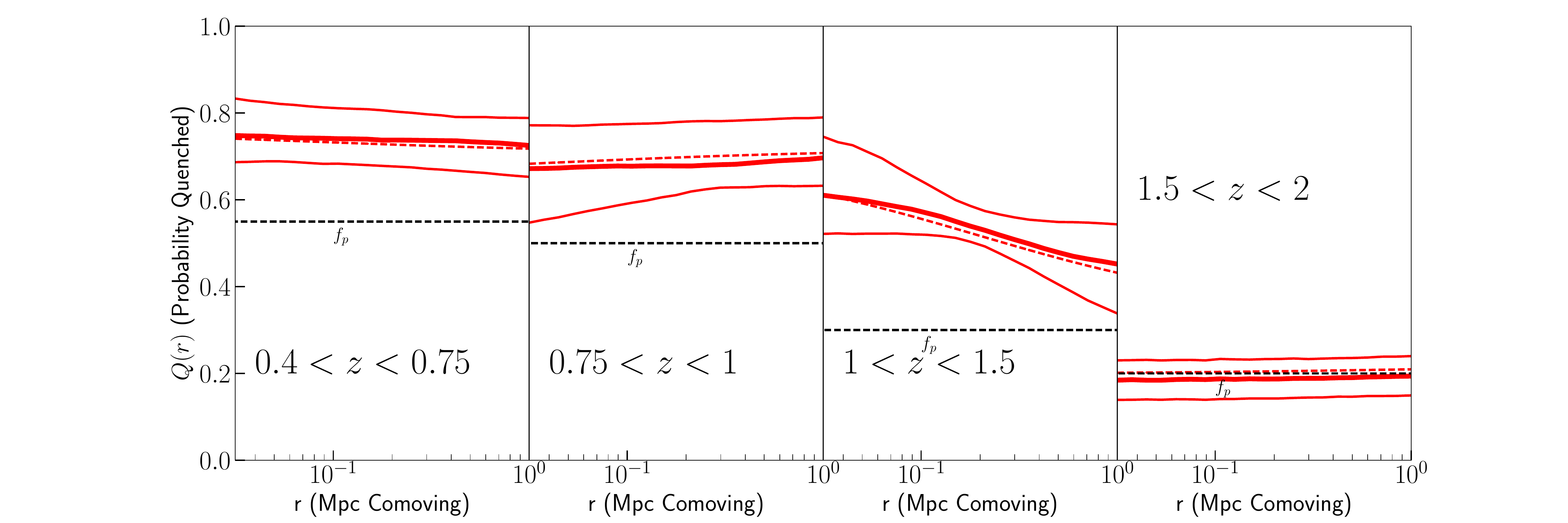}
\caption{The quenching probabilities implied by our MCMC parameter fitting for the four redshift ranges is shown in red, i.e. the probability of the lower mass galaxies being quenched as a function of radial location in halo. The full thick and two thin lines are the pointwise median and 16th and 84th percentiles respectively of our sampled models. The dashed red lines are the model corresponding to the best fit parameters. The horizontal black dashed lines are global passive fraction for the lower mass galaxies at that redshift.}
\label{fig:quench_z}
\end{figure*}

We show the data and corresponding fits for our four redshift bins in Figure~\ref{fig:fit_example}, and the triangle plot for the MCMC run of our $1<z<1.5$ redshift bin in Figure~\ref{fig:triangle} to highlight the typical posterior probabilities for each parameter. Moderate quality fits are obtained that capture the key features of the data. Figure~\ref{fig:quench_z} shows the corresponding implied probabilities of being quenched as a function of radius within the halo. The dashed lines show the global probability (the fraction of all galaxies that are passive, see table \ref{tab:counts}) of the low mass galaxies being quenched at that redshift.

At the highest redshift, the galaxies are no more likely to be quenched when inside the environment associated with the massive galaxy, than out of it. At intermediate redshifts $1<z<1.5$ there is moderate radial dependence for likelihood of being quenched - on the outskirts of the halo environment the probability of being quenched is $\sim 0.4$ (similar to the global passive fraction at this redshift), but this rises to $\sim 0.6$ in the centre of the halo. However by the lowest two redshift bins ($0.4<z<1$), the lower mass galaxies are preferentially quenched in halos hosting massive galaxies (a passive fraction of $\sim 0.75$ within the halo, versus $\sim 0.5$ outside), with very little radial dependence within the halos.

As mentioned in section \ref{sec:final_sample}, we have used global passive fractions from \citet{Darvish2016}; slightly different choices of passive fraction would lead to slightly different best fit models, but the overall qualitative behaviour would not be altered. In particular the fact that the cross-correlation functions overlap for the highest redshift bin means that the global passive fraction and the passive fraction in the massive galaxy halo must be similar for those redshifts\footnote{This depends on our model assumptions being correct (in particular the assumption that the massive tracer galaxies are centrals for all the halos of a particular mass scale). Large dispersion in halo mass or other more complex scenarios could weaken this inference.}. Furthermore the small-scale behaviour of the highest redshift cross-correlations `hints' at some small radial dependence, but the modelling does not favour it a) because of the relatively large error-bars and b) the small-scale passive low-mass to high-mass cross-correlation does not `up-turn' to match the observed `down-turn' in the star-forming low-mass to high-mass cross-correlation.

\begin{figure}
\includegraphics[scale=0.4]{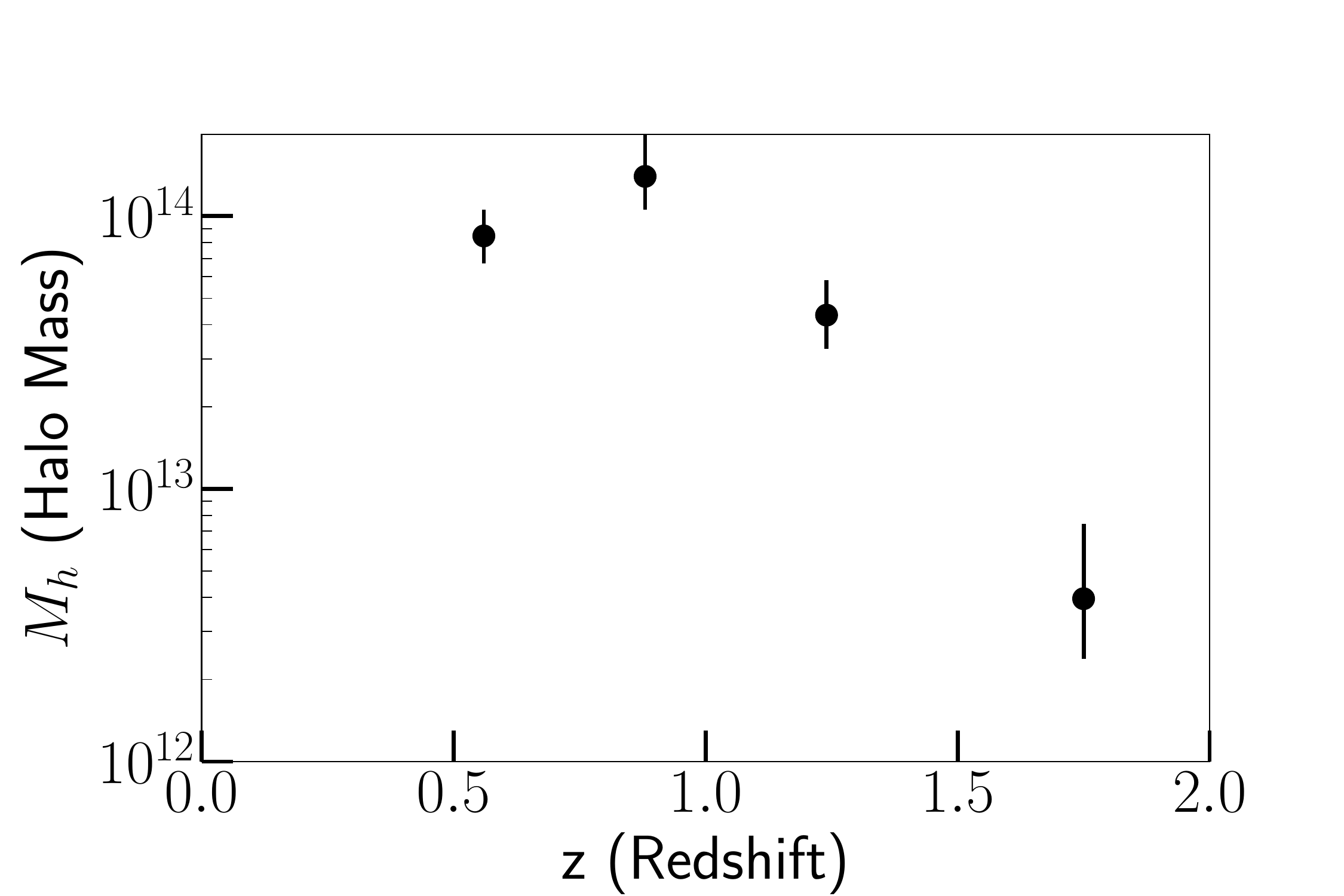}

\caption{The halo mass of the environment probed at each redshift we make measurements for. Error bars are the 16th and 84th percentiles of the MCMC fitting samples.}
\label{fig:halo_masses}
\end{figure}

We show in Fig. \ref{fig:halo_masses} the halo mass of the best fitting profile for each redshift bin, which can be seen to increase with time. This mass should not be directly interpreted as the halo mass that all the galaxies are in (typically the galaxies are found in a range of halo masses), but as the mass corresponding to the scale at which non-linear effects are needed to describe the cross-correlation functions. We interpret this increase as resulting from evolution in the HMF; at high redshift, the more massive halos are rare, and lower mass group environments are being predominantly probed as they are more common. At lower redshifts, the higher mass halos aren't as rare, and as the more massive halos have more satellites in them, they become more dominant in our measurements, meaning that non-linear effects extend to larger radii. Higher mass haloes correspond to larger radii both because a) higher mass haloes have larger virial radii, but also b) because satellite-satellite contributions start to play more of a role for higher mass haloes (which have galaxy pairs at up to twice the virial radius), as opposed to just the central-satellite contributions considered here (which only have galaxy pairs up to one virial radius). \citet{Hatfield2015} reported that $M_{min}$, the halo mass required to form a central galaxy, was $\sim 10^{12.5} M_{\sun}$ for galaxies with stellar masses corresponding to our `high mass' sample, and  $M_{1}$, the halo mass required to host satellites, to be $\sim 10^{13.7}$ for galaxies with stellar masses corresponding to our `low mass' sample. Thus our halo masses in Fig. \ref{fig:halo_masses} of $\sim 10^{12.5-14} M_{\sun}$ (virial radii of $\sim 0.4-1$Mpc  at $z=1$) seem reasonable for central-satellite interactions between galaxy samples of those two stellar mass ranges. Increasing satellite-satellite contributions towards low redshift means that our assumption that all massive galaxies are centrals becomes less accurate for our lower redshift bins, which potentially impacts on our result of little quenching radial dependence within haloes. This effect is likely small, as the satellite fraction is still $<10$ percent (\citealp{Hatfield2015}) for our massive galaxies in our lowest redshift bin, but our radial dependence measurements should still be treated with appropriate caution.

Table \ref{tab:z_fits} shows the cross biases of our measurements. Note that these are not the biases of each sample of galaxy, they are the cross biases, which is the geometric mean of the biases of the two samples being cross-correlated. Typically we would expect that at a given redshift the higher mass sample would have a higher bias than the lower mass sample, so the cross bias will be between the two values. In each redshift bin the masses of the two samples are unchanged, it is just the star formation rate of the lower mass sample that is changed.  We see a general trend of bias increasing at high redshifts, from $b\sim 1$ at $z \sim 0.6$ to $b\sim2.5$ at $z\sim2$, which is consistent with measurements in \citet{McCracken2015} and \citet{Hatfield2015}. However we see that in our highest bin, the biases are essentially the same regardless of the star formation rate of the lower mass sample, but diverge for lower redshifts, consistent with what we had qualitatively described in section \ref{sec:no_selection}. The cross-biases ought not be directly compared to the best fit $M_{h}$ values - the cross-biases depend on the biases of the two galaxy populations as a whole (dominated by central-central pairs of the high- and low-mass galaxy samples), whereas the $M_{h}$ values specifically correspond to halos that have members of both samples (dominated by central-satellite pairs of the high- and low-mass galaxy samples).

In terms of length scale of quenching $r_{Q}$ and sharpness of transition $a$, little/no constraints can be made, as typically little radial dependence for quenching probability is found. In table~\ref{tab:z_fits} one can see that typically values of around 2 are found for $a$, meaning that changes in quenching probabilities span $\sim 2$ orders or magnitudes i.e. no sharp transitions.

\begin{figure}
\includegraphics[scale=0.55]{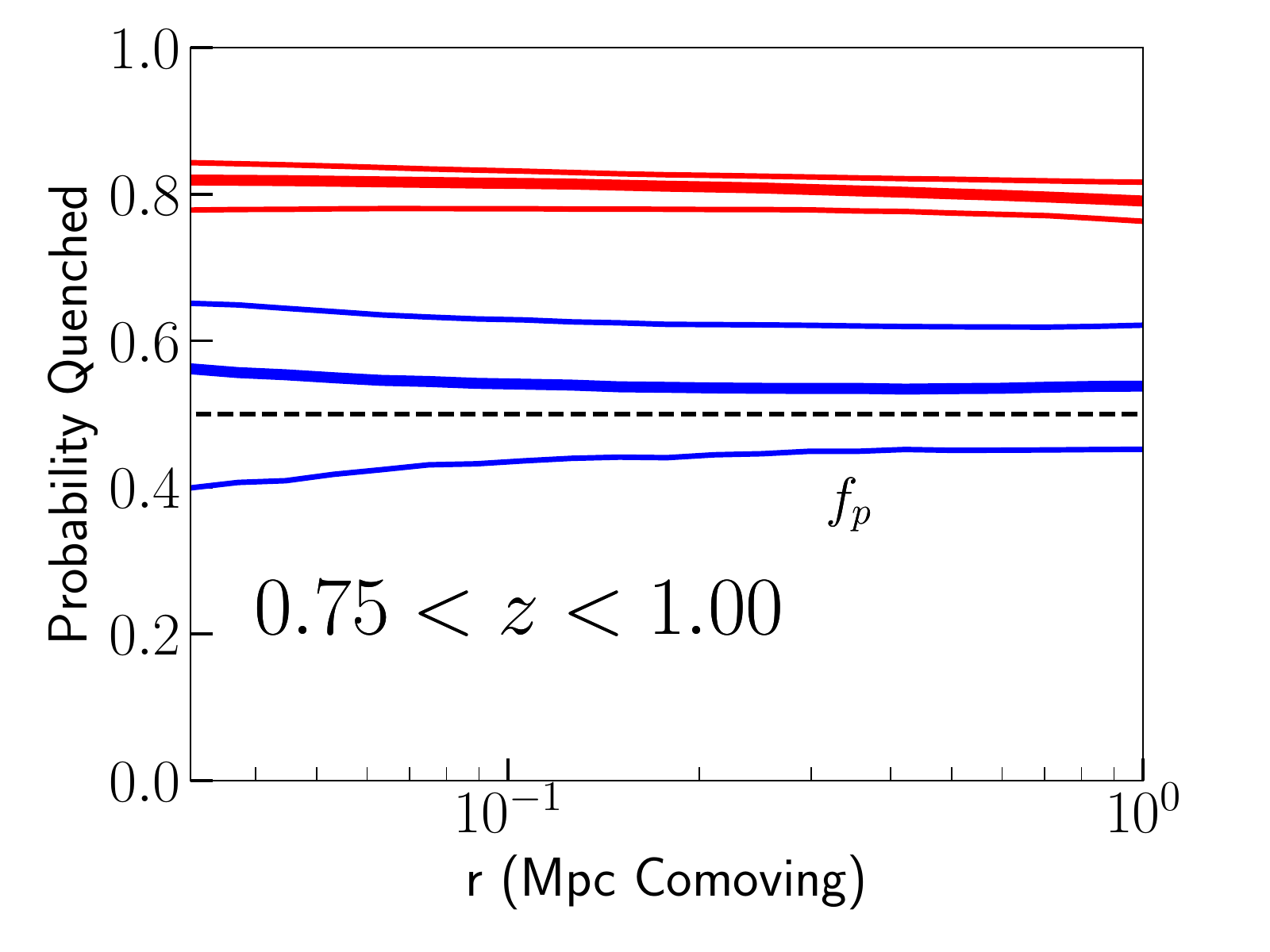}
\caption{Estimate of the probability of being quenched as a function of radial location from our best fit models when the central is passive (red), and when the central is star forming (blue). Full lines correspond to the pointwise 16th, 50th and 84th percentiles, dashed line to the quenching model of the best fitting parameters.}
\label{fig:ssfr_model_split}
\end{figure}

We also model the results of section \ref{sec:ssfr_selection} when splitting by sSFR in the massive galaxy sample, see figure \ref{fig:ssfr_model_split}. We see that galaxies are preferentially quenched when the massive galaxy is passive as opposed to star forming.

\begin{table*}
\caption {The constraints on our quenching and bias model from our MCMC results. The errors are the 16th and 84th percentiles of the posterior. The cross biases are labeled by what pairs of samples they correspond to; H-PL corresponds to the high mass sample to low mass passive sample, H-SL the high mass to low mass star forming sample.} \label{tab:z_fits}
\begin{tabular}{@{}cccccccc@{}}
\toprule
Redshift Range	& $K $	& $S$  & $r_{Q}$ & $a$ & $b_{\times}$ (H-PL) & $b_{\times}$ (H-SL) & $M_{h}$	\\
{$0.40<z<0.75$} & {$0.71\substack{+0.10 \\ -0.15}$} & {$0.76\substack{+0.11 \\ -0.12}$} & {$0.07\substack{+0.65 \\ -0.05}$} & {$1.77\substack{+0.74 \\ -0.61}$} & {$1.15\substack{+0.22 \\ -0.20}$} & {$1.04\substack{+0.25 \\ -0.22}$} & {$13.9\substack{+0.10 \\ -0.10}$}\\
{$0.75<z<1.00$} & {$0.72\substack{+0.10 \\ -0.11}$} & {$0.65\substack{+0.19 \\ -0.24}$} & {$0.03\substack{+0.10 \\ -0.02}$} & {$1.79\substack{+0.7 \\ -0.48}$} & {$1.53\substack{+0.26 \\ -0.10}$} & {$1.25\substack{+0.62 \\ -0.23}$} & {$14.1\substack{+0.18 \\ -0.12}$}\\
{$1.00<z<1.50$} & {$0.36\substack{+0.18 \\ -0.18}$} & {$0.68\substack{+0.2 \\ -0.17}$} & {$0.2\substack{+0.65 \\ -0.13}$} & {$1.36\substack{+0.68 \\ -0.69}$} & {$2.44\substack{+0.13 \\ -0.18}$} & {$2.06\substack{+0.18 \\ -0.14}$} & {$13.6\substack{+0.13 \\ -0.12}$}\\
{$1.50<z<2.00$} & {$0.23\substack{+0.13 \\ -0.19}$} & {$0.21\substack{+0.07 \\ -0.10}$} & {$1.2\substack{+2.2 \\ -1.0}$} & {$1.58\substack{+0.85 \\ -0.82}$} & {$3.41\substack{+0.04 \\ -0.10}$} & {$3.37\substack{+0.08 \\ -0.16}$} & {$12.6\substack{+0.27 \\ -0.22}$}\\
\bottomrule
\end{tabular}
\end{table*}

\section{Discussion}

\subsection{Summary of Results} \label{sec:result_summary}

Our results in sections \ref{sec:no_selection} and \ref{sec:ssfr_selection} show that targeted cross-correlation measurements (the `environment tracer' method) can be used to probe the environmental quenching and galactic conformity noted in other measurements. Modelling this within the HOD framework shows that we are probing group environments of $\sim 10^{13.5} M_{\sun}$ halos, which have a massive galaxy at the centre, and 1-2 satellites of lower masses, which agrees with \citet{Viola2015}.

The measurements made over a range of redshifts suggest a coherent `timeline' of the role of environment/the colour-density relation over the age of the Universe. The data suggests that environmental quenching is not a significant factor for the galaxies under consideration in this study at $z \sim 2$, if it was there would be differences in the clustering that depended on the star formation rates of the galaxies. In the $1<z< 1.5$ bin, environmental quenching has started to play a role on scales smaller than $\sim 1$Mpc, and by $z \sim 0.5$ environmental quenching has played such a part that passive galaxies are found in dramatically different environments to star forming galaxies of the same mass and have different biases etc.

We included in our model the probability of a galaxy being quenched having a radial dependence within the halo; however for the most part we found that the data favoured a flat probability at lower redshift, with moderate radial dependence at $1<z<1.5$. Our radial dependence measurements at low redshift should be appropriately caveated, as discussed in section \ref{sec:quenching_model_results}. There are several possibilities for why this may be. It could be that the motions of the satellites within the halo quickly smear out any signal of spatial dependence of quenching e.g. satellites are preferentially quenched in the centre of halos, but then quickly move out again, weakening the signal. Alternatively it could be possible that quenching is associated with time spent in a halo; \citet{Smethurst2015} finds that galaxies closer to the centre of a group halo were quenched earlier. This time dependence, combined with the fact that sub-halos accreted earlier are typically found towards the centre of the halo\footnote{It appears to be the case that sub-haloes are distributed independently of sub-halo mass, but with some accretion time dependence, within the larger halo, but see \citet{Han2015} for a more detailed discussion.}, could mean that at $z \sim 0$ a radial dependence is observed as there is a large range of accretion times of sub-halos, but at the comparatively early times probed in this study there was a much smaller range of accretion times for sub-halos, limiting the possible radial dependence.

Our results show that the satellite quenched fraction is lower when the massive galaxy is star forming (although still higher than the global fraction). We interpret this as a measurement of how 1-halo conformity can manifest itself in the correlation function, although we note that this does not perfectly control for halo mass (as the passive and star-forming massive galaxies are likely centrals in haloes of slightly different masses - see section \ref{sec:future} for a discussion of how this could be controlled in a full HOD with quenching model). Galaxies are preferentially quenched as satellites when the central galaxy is quenched, and vice versa. We also see 2-halo conformity in our results in the sense that on scales larger than individual halos, the difference between the passive/star-forming low-mass cross-correlation amplitudes is greater when the massive galaxy is passive than when star-forming. Note we do not attempt to distinguish between the weaker sense of 2-halo conformity (correlations that are a natural result of the fact that massive halos are typically near other massive halos, and massive objects typically host more quenched objects) and the stronger claim that there is an environmental effect over volumes of space much bigger than halos that can have an effect on galaxies.  \citet{Hearin2015} note that 1-halo conformity and 2-halo conformity are not necessarily separate effects, as satellites lived most of their lives as nearby central galaxies before accretion. Our results are consistent with the view presented using SDSS data at $z\sim0$ in \citet{Knobel2014,Paranjape2015,Pahwa2017} that conformity is strong within haloes, but that correlations on scales larger than halo scales are simply a result of halo bias.  \citet{Knobel2014} use the groups and halo masses of \citet{Yang2011}, and find that satellite galaxy sSFR is correlated with the sSFR of its central galaxy, at fixed halo mass, central and satellite stellar mass, distance from the centre of halo, and large-scale overdensity. \citet{Paranjape2015} show by generating mock catalogues that halo-halo correlations are not required to generate a signal on scales larger than individual haloes. \citet{Pahwa2017}, similarly to our work, develop an approach to modelling the effects of conformity on clustering measurements using a modification of HOD schemes. They find using SDSS clustering measurements similar levels of 1-halo conformity to \citet{Knobel2014}, with no evidence for halo-halo correlations. \citet{Zu2015,Zu2015a} also provide a highly relevant evolution of the HOD model, fitting to clustering and galaxy-galaxy lensing in SDSS, finding that models with halo mass as the sole driver of probability of being quenched fit the data best.  \citet{Tinker2017} also find that halo-halo correlations are negligible using halo age matching models. At higher redshift, \citet{Kawinwanichakij2015} investigate conformity at $0.3<z<2.3$, finding that conformity is most significant at to $0.6<z < 1.6$, and \citet{Berti2016} investigate $0.2<z<1$, finding strong evidence for substantial conformity on scales $<1$Mpc, and marginal evidence for a small amount of conformity on scales $1-3$Mpc. There does certainly seem to be some dependence on the large-scale density field in some contexts (e.g. the stellar to halo mass ratio is found to have a slight dependence on it in \citealp{Cen2015}), but in summary most of the literature seems to find that conformity is only relevant on the scales of individual haloes, but can be significant up to $z \sim 1.5$.

\subsection{Quenching over cosmic time} \label{sec:lit_compare}

Direct comparisons of measurements like `passive fraction' and `red fraction' in the literature are difficult a) because these measurements are functions of population, so comparisons are only direct between matched stellar mass ranges and redshifts etc. b) these fractions are dependent on choice of dividing colour or sSFR break and c) photometric sSFR values remain tentative, and can vary with what photometry is available and choice of templates etc. Nonetheless it is possible to see whether the broad picture suggested by our results is consistent with other studies.

Observationally, \citet{Woo2012} used the SDSS ($z \sim 0.1$) to investigate the role of halo mass and location in halo on quenching. Their key findings were that environmental effects on quenching seemed more correlated with halo mass at fixed stellar mass, than with stellar mass at fixed halo mass, suggesting that halo mass was the more important factor. They found that quenched fraction increased with halo mass, and increased towards the centre of the halo, most notably in the central $0.1-0.3 R_{vir}$. This appears to be in contradiction to our results, which exhibit little radial dependence, but closer inspection shows that at $\sim 10^{13} M_{\sun}$, the halo mass we are predominantly probing here, they find the quenched fraction is flat at about 0.5 for the inner part of the halo, and drops off to 0.2 at around $0.6 R_{vir}$, so future work is needed to see if clustering recovers the radial dependence they measure. In the GAMA survey \citet{Prescott2011} find similar results, of quenched satellite fraction increasing with the mass of the central galaxy, with little radial dependence at lower masses, and some radial dependence on small scales for the more massive galaxies. In agreement with our results/1-halo conformity, they also find that the quenched fraction is higher when the central is passive (see their figure 7).

In terms of simulations, \citet{Gabor2014} discuss a cosmological simulation with a prescription for environmental quenching within halos in areas dominated by hot gas. A key finding is that a lot of what might be interpreted as `environment' quenching is really `mass quenching' in `pre-processed' satellites; satellites at $z=0$ used to be centrals at higher redshift and experienced mass quenching before accreting and becoming a satellite. In addition they find that, although with larger scatter, galaxies closer to the centre accreted earlier i.e. galaxies towards the edge of the halo have been there for a shorter period of time, suggesting there is also a radial component to how long galaxies have been in the environment. Similarly a fraction of galaxies are `ejected' so it is possible to have galaxies that get environmentally quenched, but then are found far away from that environment. However they find that there is a strong dependence on the quenched fraction on both radial location in halo, and halo mass (with their results agreeing with \citealp{Woo2012}). They find a quenched fraction of $\sim 0.3$ everywhere in the halo for halo masses $\sim 10^{12} M_{\sun}$, spanning 0.3 at the edge of the halo to near unity at the centre of the halo for halo masses $\sim 10^{13} M_{\sun}$, and nearly all quenched except for the very edge of the halo for $> 10^{13} M_{\sun}$. This appears to have slightly more radial dependence than we find, but is at $z=0$, so isn't directly comparable to our measurements. In addition, at the halo masses comparable to our results, the quenching probability is actually quite flat apart from a) the very centre of the halo ($<0.05 R_{vir}$ and b) beyond the virial radius, which we can't account for within our formalism, so it is not clear if our results of little radial dependence are in tension or not.

Both \citet{Woo2012} and \citet{Gabor2014} are at $z \sim 0$. In terms of $z \sim 1$, there is substantial disagreement - some authors claim that the sSFR-density relation is qualitatively the same as $z \sim 0$ relation, while other authors have suggested that the sSFR-density relation actually essentially swaps at this epoch e.g. sSFRs are higher in denser environments, in contrast to the local universe. \citet{Patel2009} and \citet{Baldry2006} both report the $z \sim 0$ relation holding at $z \sim 1$ in Spitzer and Gemini data respectively. \citet{Chuter2011} and \citet{Quadri2012} find that the relation holds up to $z \sim 2$ in UDS data. \citet{Duivenvoorden2016} find in the Herschel Extragalactic Legacy Survey (HELP) that their data is consistent with no environmental dependence at $z>2$. Conversely, \citet{Elbaz2007} and \citet{Welikala2015}, find the relation reversing in the Great Observatories Origins Deep Survey and South Pole Telescope data (with some simulations reporting similar results e.g.  \citealp{Tonnesen2014}). Our results are in general not supportive of the relation reversing (although more work is needed to confirm that environments and host galaxies etc. are being compared consistently) - the picture we see in our data is that of the relation weakening, until it only holds on small scales by $z \sim 1.5$, and disappearing by $z \sim 2$, which is the trend the weight of the literature appears to be behind. If the relation reversed, by higher redshifts one might expect to see the star forming lower mass galaxies being more clustered around our tracers of dense environments, the massive galaxies, and one might expect the biases of star forming galaxies to be higher than that of passive galaxies of the same mass at higher redshifts, which we do not see. However it will take more work on the development of incorporating quenching mechanisms into the HOD/clustering framework before we can be sure we are consistently comparing the same galaxies in the same environments between different probes of density.

With regards to understanding why it might be that the sSFR-density relation emerges only at later times, \citet{Peng2010a} offers some possible insight. They study environment in SDSS and zCOSMOS up to $z \sim 1$, looking at red/blueness in relation to local overdensity $\delta$, and explicitly attempt to empirically model the measurements without the invocation of physical models e.g. the idea that galaxies live and are formed in halos. They find, among other things, that the differential effect of a given environment doesn't change with epoch, but that an environmental dependence of $f_{red}$ emerges over time, because galaxies move to higher overdensities as large scale structure grows in the Universe, not because there is any change over time of the effect of a given environment. They have a relative quenching efficiency, which characterises at fixed stellar mass how much more quenched galaxies are in some environment compared with some reference environment. This is in broad agreement with the picture suggested by our results, of the environment not yet having had enough time to have had effect at high redshift, and only coming into significance later than $z \sim 1$. For understanding why there appears to be little radial dependence to the probability of quenching, \citet{McGee2014} discuss an updated view of the baryon cycle and the quenching of accreted satellite galaxies in halos. One paradigm of satellite quenching is that galaxies that are accreted are forming stars from a reservoir of gas, that is slowly removed from the stellar disk in the case of ram pressure stripping, or from the host sub-halo in the case of strangulation. However \citet{McGee2014} discuss observations and simulations that suggest a slightly different view - that star formation in field galaxies is driven by fresh inflows, as opposed to feeding on a reservoir. In this scenario, quenching of galaxies when they are accreted by a larger halo is a result of these inflows stopping, as opposed to gas from a reservoir being removed by any sort of stripping. \citet{McGee2014} discuss evidence that these inflows are stopped when the accreted galaxy passes $\sim 1.8 R_{vir}$, which could explain our lack of radial dependence; quenching merely depends on passing that radius, as opposed to some radius dependent stripping process.

With regards to the use of correlation functions to make inferences about environmental effects, \citet{Skibba2009} \textit{marked} correlation function statistics using SDSS Galaxy Zoo data, where essentially the conventional correlation function is compared with a weighted correlation function e.g. galaxies could be weighted by their colour etc. Given that, as well as the colour-density relation, there is a morphology-density relation, it is natural to ask which came first e.g. is colour fundamentally tied to density, and the morphology-density relation just arrises because colour and morphology are correlated, or vice versa, or both/neither? They found that, at fixed colour, a correlation function marked by morphology had no enhancement. However, at fixed morphology (and luminosity), a correlation function marked by colour had substantial enhancement - suggesting that the colour-density relation is the more fundamental relation. This agrees with our results that correlation function measurements can give meaningful insight into the colour-density relation, as well as being reassuring that colour-density is the more fundamental relation to investigate (seeing as ground based surveys in general have comparatively little morphological information at high redshift).

Our results in terms of environmental dependence appearing in the 1-halo term agree well with \citet{Hearin2015}, who suggest that there should be little to no conformity at high redshift, 1-halo conformity at intermediate redshifts, and be present in both 1-halo and 2-halo scales at redshifts towards $z\sim0$. This is very similar to the picture we see of no enhancement at high redshift, enhancement on small scales at intermediate redshifts, and enhancement on all scales at the lowest redshifts.

\citet{Darvish2016} and \citet{Hartley2014a} are perhaps the most similar study to ours. \citet{Darvish2016} look at galaxies with photometric redshifts up to $z \sim 3$ in 2 sq deg of the COSMOS field, with a limiting $K$-band magnitude of 24. The use Voroni tessellation to give their density measure (in the manner of \citealp{Darvish2015}), and find that at fixed stellar mass sSFRs are up to two orders of magnitude lower in denser regions, but that at higher redshifts ($z>1$) this dependence essentially vanishes. This is in qualitative agreement with our result that the sSFR-density relation predominantly comes into action at lower redshifts. \citet{Hartley2014a} (who implicitly also cross-correlate high and low mass galaxies) investigate radial profiles of satellite number densities around centrals for very similar stellar mass and redshift ranges to our work, also finding that passive satellites are preferentially located around passive centrals, and that the quenched fraction around star-forming galaxies was similar to the field quenched fraction. They found comparatively very little evidence for evolution, finding that environmental trends were still in place at $z \sim 2$. Future data sets (e.g. the full 12 deg$^2$ of VIDEO) should be able to more definitively confirm whether or not the role of environment changes or not over cosmic time.

\subsection{Future Development of the Approach} \label{sec:future}

As discussed in section \ref{sec:result_summary}, the modelling discussed here relies on some approximations that start to fail in some regimes probed in this paper. In addition there are still degeneracies in the modelling, and Scenario 3 is less revealing than Scenario 1 or 2 as it does not allow discrimination between quenching from location in the halo, and quenching from proximity to galaxy. Both of these mean that a logical next step is to incorporate a quenching model for all halo masses and stellar masses, so that both central-satellites and satellite-satellite pairs can be incorporated, from all halo masses. As more data become available, and the model is developed, we believe that a more robust understanding of interactions between galaxies within a halo will be obtainable, and more definitive statements about the distribution of star formation in galaxies within a halo will become possible.

More complex scenarios, where the quenching mechanism is more complicated  than considered here can be probed by simulating galaxy locations in halos, as we do for the auto-correlations of Scenario 1. The massive galaxy and less massive galaxy have locations in the halo drawn from a distribution that traces the halo profile. Then according to equation \ref{eq:quenching_formula}, the lower mass galaxy is determined probabilistically to be passive or star-forming. This allows us to construct essentially $D_{1}D_{2}$ of the correlation function, and therefore a 1-halo term corresponding to a physical quenching process. This can be used to measure the effects of multiple quenching sources in the same halo etc.

In general, for a model that specified galaxy occupation numbers and interaction terms, all possible auto and cross correlation 1-halo terms can be constructed in this Monte Carlo manner: 1) sample halo mass from the HMF 2) sample how many galaxies are in the halo (given an average occupation number as a function of halo mass and assuming Poisson) 3) sample locations for the galaxies within the halo 4) determine star forming/AGN/galaxy properties of choice for galaxies based on probabilistic model of galaxy interactions 5) record the number of galaxies of each species and all pairs of spatial separation. This allows one to construct all auto and cross 1-halo terms for arbitrarily complicated models of the halo environment. The parameters that describe the halo environment interactions can then be explored with MCMC techniques. This approach is much more computationally challenging than the cases for which analytic expressions are available, but may provide a useful intermediate between the cases described by analytic expressions, and a full-blown cosmological simulation. Another possible approach would be to `graft' a galaxy occupation and interaction model on a dark matter only simulation as per the comprehensive {\sc Halotools}  code described in \citet{Hearin2016}.

Once the physical model is used to construct the 1-halo term, it can be added to a 2-halo term, simply constructed by scaling the dark matter correlation function by the galaxy biases of the two samples, as discussed in section \ref{sec:ACF_def}. This spatial cross-correlation function can then be compared with observations by projecting to an angular correlation function etc.

We note that understanding the cross-correlation function is important for other applications e.g. \citet{Croft2013} consider a very similar scenario to us of high-mass galaxies cross-correlated with low-mass galaxies, for the purpose of detecting gravitational redshift-space distortions.

\section{Conclusions} \label{sec:conclusions}

We have used data from the VIDEO survey to investigate the processes which results in environmental quenching and galactic conformity. In particular we have studied the cross-clustering of galaxies with the two point cross-correlation function up to $z\sim2$, varying the stellar mass and star formation rates of the galaxies. Building on the JHOD formalism introduced by \citet{Simon2008} we used MCMC methods to explore the probability of a galaxy being quenched at different redshifts, different radial locations within a halo, and when the central galaxy was star forming or passive.

Key conclusions:
\begin{itemize}
\item Cross correlations between galaxy samples of different star formation rates contain information about the role of environment in determining whether a galaxy is star forming or passive
\item Measurements of the cross-correlations suggest that the sSFR-density relation is non-existent/weak at $z \sim 1.75$
\item The sSFR-density/environmental quenching can be seen to emerge at $z \sim 1$ and grows increasingly important towards $z=0$ as galaxies are increasingly more likely to be quenched in group environments over the field
\item Lower mass galaxies appear to be more likely to be quenched in the vicinity of a massive passive galaxy, than a massive star forming galaxy of the same stellar mass
\item There appears to be little radial dependence for being quenched within a halo at lower redshifts, with some evidence for moderate dependancy at $z\sim1.25$, supportive of the quenching of accreted galaxies being caused by the cessation of fresh inflows of gas, as opposed to tidal stripping
\item A quenching mechanism within a halo can be added into the HOD framework, and may be a promising approach for future environment studies, as well as being important to understand for advancing the HOD programme of understanding the galaxy-halo connection
\end{itemize}

Planned future work will develop the formalism so that all halo masses are incorporated and properly controlled for. We suggest that in the future, this more sophisticated model, combined with bayesian model selection methods may allow one to discriminate between different forms of environmental quenching e.g. whether environmental quenching is really associated with halo mass, location in the halo, proximity to another galaxy or the presence of an active galactic nucleus.

\section*{Acknowledgements}

The authors thank Will Hartley for useful comments that have greatly improved this paper.

PWH wishes to acknowledge support provided through an STFC studentship, and the Rector and Fellows of Lincoln College for support through the Graduate Research Fund. MJJ acknowledges support from the UK Science and Technology Facilities Council [ST/N000919/1].

Based on data products from observations made with ESO Telescopes at the La Silla or Paranal Observatories under ESO programme ID 179.A- 2006.

Based on observations obtained with MegaPrime/MegaCam, a joint project of CFHT and CEA/IRFU, at the Canada-France-Hawaii Telescope (CFHT) which is operated by the National Research Council (NRC) of Canada, the Institut National des Science de l'Univers of the Centre National de la Recherche Scientifique (CNRS) of France, and the University of Hawaii. This work is based in part on data products produced at Terapix available at the Canadian Astronomy Data Centre as part of the Canada-France-Hawaii Telescope Legacy Survey, a collaborative project of NRC and CNRS.

\bibliographystyle{mn2e_mod}
\bibliography{conformity_paper}

\begin{thebibliography}{88}
\expandafter\ifx\csname natexlab\endcsname\relax\def\natexlab#1{#1}\fi

\bibitem[{Arnouts {et~al}\mbox{.}(1999)Arnouts, Cristiani, Moscardini,
  Matarrese, Lucchin, Fontana, \& Giallongo}]{Arnouts1999}
Arnouts S., Cristiani S., Moscardini L., Matarrese S., Lucchin F., Fontana A.,
  Giallongo E., 1999, Monthly Notices of the Royal Astronomical Society, 310,
  540

\bibitem[{Arnouts {et~al}\mbox{.}(2002)Arnouts, Moscardini, Vanzella, Colombi,
  Cristiani, Fontana, Giallongo, Matarrese, \& Saracco}]{Arnouts2002}
Arnouts S. {et~al.}, 2002, Monthly Notices of the Royal Astronomical Society,
  329, 355

\bibitem[{Asorey {et~al}\mbox{.}(2016)Asorey, {Carrasco Kind}, Sevilla-Noarbe,
  Brunner, \& Thaler}]{Asorey2016a}
Asorey J., {Carrasco Kind} M., Sevilla-Noarbe I., Brunner R.~J., Thaler J.,
  2016, Monthly Notices of the Royal Astronomical Society, 459, 1293

\bibitem[{Baldry {et~al}\mbox{.}(2006)Baldry, Balogh, Bower, Glazebrook,
  Nichol, Bamford, \& Budavari}]{Baldry2006}
Baldry I.~K., Balogh M.~L., Bower R.~G., Glazebrook K., Nichol R.~C., Bamford
  S.~P., Budavari T., 2006, Monthly Notices of the Royal Astronomical Society,
  373, 469

\bibitem[{Baldry {et~al}\mbox{.}(2010)Baldry, Robotham, Hill, Driver, Liske,
  Norberg, Bamford, Hopkins, Loveday, Peacock, Cameron, Croom, Cross, Doyle,
  Dye, Frenk, Jones, van Kampen, Kelvin, Nichol, Parkinson, Popescu, Prescott,
  Sharp, Sutherland, Thomas, \& Tuffs}]{Baldry2010}
Baldry I.~K. {et~al.}, 2010, Monthly Notices of the Royal Astronomical Society,
  404, 86

\bibitem[{Ball {et~al}\mbox{.}(2007)Ball, Loveday, \& Brunner}]{Ball2006}
Ball N.~M., Loveday J., Brunner R.~J., 2007, Monthly Notices of the Royal
  Astronomical Society, 383, 907

\bibitem[{Balogh {et~al}\mbox{.}(2004)Balogh, Eke, Miller, Lewis, Bower, Couch,
  Nichol, Bland-Hawthorn, Baldry, Baugh, Bridges, Cannon, Cole, Colless,
  Collins, Cross, Dalton, Propris, Driver, Efstathiou, Ellis, Frenk,
  Glazebrook, Gomez, Gray, Hawkins, Jackson, Lahav, Lumsden, Maddox, Madgwick,
  Norberg, Peacock, Percival, Peterson, Sutherland, \& Taylor}]{Balogh2004}
Balogh M. {et~al.}, 2004, Monthly Notices of the Royal Astronomical Society,
  348, 1355

\bibitem[{Bell {et~al}\mbox{.}(2004)Bell, Wolf, Meisenheimer, Rix, Borch, Dye,
  Kleinheinrich, Wisotzki, \& McIntosh}]{Bell2004}
Bell E.~F. {et~al.}, 2004, The Astrophysical Journal, 608, 752

\bibitem[{Berti {et~al}\mbox{.}(2017)Berti, Coil, Behroozi, Eisenstein, Bray,
  Cool, \& Moustakas}]{Berti2016}
Berti A.~M., Coil A.~L., Behroozi P.~S., Eisenstein D.~J., Bray A.~D., Cool
  R.~J., Moustakas J., 2017, The Astrophysical Journal, 834, 87

\bibitem[{Bertin \& Arnouts(1996)}]{Bertin1996}
Bertin E., Arnouts S., 1996, Astronomy and Astrophysics Supplement Series, 117,
  393

\bibitem[{Beutler {et~al}\mbox{.}(2011)Beutler, Blake, Colless, Jones,
  Staveley-Smith, Campbell, Parker, Saunders, \& Watson}]{Beutler2011}
Beutler F. {et~al.}, 2011, Monthly Notices of the Royal Astronomical Society,
  416, 3017

\bibitem[{Bruzual \& Charlot(2003)}]{Bruzual2003b}
Bruzual G., Charlot S., 2003, Monthly Notices of the Royal Astronomical
  Society, 344, 1000

\bibitem[{Cen \& Safarzadeh(2014)}]{Cen2015}
Cen R., Safarzadeh M., 2014, The Astrophysical Journal, 798, L38

\bibitem[{Chiu {et~al}\mbox{.}(2016)Chiu, Saro, Mohr, Desai, Bocquet, Capasso,
  Gangkofner, Gupta, \& Liu}]{Chiu2016}
Chiu I. {et~al.}, 2016, Monthly Notices of the Royal Astronomical Society, 458,
  379

\bibitem[{Chuter {et~al}\mbox{.}(2011)Chuter, Almaini, Hartley, McLure, Dunlop,
  Foucaud, Conselice, Simpson, Cirasuolo, \& Bradshaw}]{Chuter2011}
Chuter R.~W. {et~al.}, 2011, Monthly Notices of the Royal Astronomical Society,
  413, 1678

\bibitem[{Coe(2010)}]{Coe2010}
Coe D., 2010, eprint arXiv:1005.0411

\bibitem[{Conselice(2014)}]{Conselice2014}
Conselice C.~J., 2014, Annual Review of Astronomy and Astrophysics, 52, 291

\bibitem[{Cooray \& Sheth(2002)}]{Cooray2002}
Cooray A., Sheth R., 2002, Physics Reports, 372, 1

\bibitem[{Coupon {et~al}\mbox{.}(2015)Coupon, Arnouts, van Waerbeke, Moutard,
  Ilbert, van Uitert, Erben, Garilli, Guzzo, Heymans, Hildebrandt, Hoekstra,
  Kilbinger, Kitching, Mellier, Miller, Scodeggio, Bonnett, Branchini,
  Davidzon, {De Lucia}, Fritz, Fu, Hudelot, Hudson, Kuijken, Leauthaud, {Le
  Fevre}, McCracken, Moscardini, Rowe, Schrabback, Semboloni, \&
  Velander}]{Coupon2015}
Coupon J. {et~al.}, 2015, Monthly Notices of the Royal Astronomical Society,
  449, 1352

\bibitem[{Coupon {et~al}\mbox{.}(2012)Coupon, Kilbinger, McCracken, Ilbert,
  Arnouts, Mellier, Abbas, de~la Torre, Goranova, Hudelot, Kneib, \& {Le
  F{\`{e}}vre}}]{Coupon2012}
Coupon J. {et~al.}, 2012, Astronomy {\&} Astrophysics, 542, A5

\bibitem[{Croft(2013)}]{Croft2013}
Croft R. A.~C., 2013, Monthly Notices of the Royal Astronomical Society, 434,
  3008

\bibitem[{Darvish {et~al}\mbox{.}(2017)Darvish, Mobasher, Martin, Sobral,
  Scoville, Stroe, Hemmati, \& Kartaltepe}]{Darvish2017}
Darvish B., Mobasher B., Martin D.~C., Sobral D., Scoville N., Stroe A.,
  Hemmati S., Kartaltepe J., 2017, The Astrophysical Journal, 837, 16

\bibitem[{Darvish {et~al}\mbox{.}(2016)Darvish, Mobasher, Sobral, Rettura,
  Scoville, Faisst, \& Capak}]{Darvish2016}
Darvish B., Mobasher B., Sobral D., Rettura A., Scoville N., Faisst A., Capak
  P., 2016, The Astrophysical Journal, 825, 113

\bibitem[{Darvish {et~al}\mbox{.}(2015)Darvish, Mobasher, Sobral, Scoville, \&
  Aragon-Calvo}]{Darvish2015}
Darvish B., Mobasher B., Sobral D., Scoville N., Aragon-Calvo M., 2015, The
  Astrophysical Journal, 805, 121

\bibitem[{Davis \& Geller(1976)}]{Davis1976}
Davis M., Geller M.~J., 1976, The Astrophysical Journal, 208, 13

\bibitem[{Davis \& Peebles(1983)}]{Davis1983}
Davis M., Peebles P. J.~E., 1983, The Astrophysical Journal, 267, 465

\bibitem[{Dressler(1980)}]{Dressler1980}
Dressler A., 1980, The Astrophysical Journal, 236, 351

\bibitem[{Duivenvoorden {et~al}\mbox{.}(2016)Duivenvoorden, Oliver, Buat,
  Darvish, Efstathiou, Farrah, Griffin, Hurley, Ibar, Jarvis, Papadopoulos,
  Sargent, Scott, Scudder, Symeonidis, Vaccari, Viero, \&
  Wang}]{Duivenvoorden2016}
Duivenvoorden S. {et~al.}, 2016, Monthly Notices of the Royal Astronomical
  Society, 462, 277

\bibitem[{Elbaz {et~al}\mbox{.}(2007)Elbaz, Daddi, {Le Borgne}, Dickinson,
  Alexander, Chary, Starck, Brandt, Kitzbichler, MacDonald, Nonino, Popesso,
  Stern, \& Vanzella}]{Elbaz2007}
Elbaz D. {et~al.}, 2007, Astronomy and Astrophysics, 468, 33

\bibitem[{Fine {et~al}\mbox{.}(2015)Fine, Shanks, Johnston, Jarvis, \&
  Mauch}]{Fine2015}
Fine S., Shanks T., Johnston R., Jarvis M.~J., Mauch T., 2015, Monthly Notices
  of the Royal Astronomical Society, 452, 2692

\bibitem[{Fisher {et~al}\mbox{.}(1994)Fisher, Davis, Strauss, Yahil, \&
  Huchra}]{Fisher1994}
Fisher J., Davis K.~B., Strauss M., Yahil M.~A., Huchra A., 1994, Monthly
  Notices of the Royal Astronomical Society, 266

\bibitem[{Foreman-Mackey {et~al}\mbox{.}(2013)Foreman-Mackey, Hogg, Lang, \&
  Goodman}]{Foreman-Mackey2012}
Foreman-Mackey D., Hogg D.~W., Lang D., Goodman J., 2013, Publications of the
  Astronomical Society of the Pacific, 125, 306

\bibitem[{Fossati {et~al}\mbox{.}(2017)Fossati, Wilman, Mendel, Saglia,
  Galametz, Beifiori, Bender, Chan, Fabricius, Bandara, Brammer, Davies,
  {F{\"{o}}rster Schreiber}, Genzel, Hartley, Kulkarni, Lang, Momcheva, Nelson,
  Skelton, Tacconi, Tadaki, {\"{U}}bler, van Dokkum, Wisnioski, Whitaker,
  Wuyts, \& Wuyts}]{Fossati2016}
Fossati M. {et~al.}, 2017, The Astrophysical Journal, 835, 153

\bibitem[{Gabor \& Dave(2014)}]{Gabor2014}
Gabor J.~M., Dave R., 2014, Monthly Notices of the Royal Astronomical Society,
  447, 374

\bibitem[{Gwyn(2012)}]{Gwyn2012}
Gwyn S. D.~J., 2012, The Astronomical Journal, 143, 38

\bibitem[{Han {et~al}\mbox{.}(2016)Han, Cole, Frenk, \& Jing}]{Han2015}
Han J., Cole S., Frenk C.~S., Jing Y., 2016, Monthly Notices of the Royal
  Astronomical Society, 457, 1208

\bibitem[{Hartley {et~al}\mbox{.}(2013)Hartley, Almaini, Mortlock, Conselice,
  Grutzbauch, Simpson, Bradshaw, Chuter, Foucaud, Cirasuolo, Dunlop, McLure, \&
  Pearce}]{Hartley2013a}
Hartley W.~G. {et~al.}, 2013, Monthly Notices of the Royal Astronomical
  Society, 431, 3045

\bibitem[{Hartley {et~al}\mbox{.}(2015)Hartley, Conselice, Mortlock, Foucaud,
  \& Simpson}]{Hartley2014a}
Hartley W.~G., Conselice C.~J., Mortlock A., Foucaud S., Simpson C., 2015,
  Monthly Notices of the Royal Astronomical Society, 451, 1613

\bibitem[{Hatfield {et~al}\mbox{.}(2016)Hatfield, Lindsay, Jarvis,
  H{\"{a}}u{\ss}ler, Vaccari, \& Verma}]{Hatfield2015}
Hatfield P.~W., Lindsay S.~N., Jarvis M.~J., H{\"{a}}u{\ss}ler B., Vaccari M.,
  Verma A., 2016, Monthly Notices of the Royal Astronomical Society, 459, 2618

\bibitem[{Hearin {et~al}\mbox{.}(2016{\natexlab{a}})Hearin, Behroozi, \&
  van~den Bosch}]{Hearin2015}
Hearin A.~P., Behroozi P.~S., van~den Bosch F.~C., 2016{\natexlab{a}}, Monthly
  Notices of the Royal Astronomical Society, 461, 2135

\bibitem[{Hearin {et~al}\mbox{.}(2016{\natexlab{b}})Hearin, Zentner, van~den
  Bosch, Campbell, \& Tollerud}]{Hearin2016}
Hearin A.~P., Zentner A.~R., van~den Bosch F.~C., Campbell D., Tollerud E.,
  2016{\natexlab{b}}, Monthly Notices of the Royal Astronomical Society, 460,
  2552

\bibitem[{Hirschmann {et~al}\mbox{.}(2014)Hirschmann, {De Lucia}, Wilman,
  Weinmann, Iovino, Cucciati, Zibetti, \& Villalobos}]{Hirschmann2014}
Hirschmann M., {De Lucia} G., Wilman D., Weinmann S., Iovino A., Cucciati O.,
  Zibetti S., Villalobos A., 2014, Monthly Notices of the Royal Astronomical
  Society, 444, 2938

\bibitem[{Ilbert {et~al}\mbox{.}(2015)Ilbert, Arnouts, {Le Floc'h}, Aussel,
  Bethermin, Capak, Hsieh, Kajisawa, Karim, {Le F{\`{e}}vre}, Lee, Lilly,
  McCracken, Michel-Dansac, Moutard, Renzini, Salvato, Sanders, Scoville,
  Sheth, Silverman, Smol{\v{c}}i{\'{c}}, Taniguchi, \& Tresse}]{Ilbert2015}
Ilbert O. {et~al.}, 2015, Astronomy {\&} Astrophysics, 579, A2

\bibitem[{Ilbert {et~al}\mbox{.}(2006)Ilbert, Arnouts, McCracken, Bolzonella,
  Bertin, {Le F{\`{e}}vre}, Mellier, Zamorani, Pell{\`{o}}, Iovino, Tresse, {Le
  Brun}, Bottini, Garilli, Maccagni, Picat, Scaramella, Scodeggio, Vettolani,
  Zanichelli, Adami, Bardelli, Cappi, Charlot, Ciliegi, Contini, Cucciati,
  Foucaud, Franzetti, Gavignaud, Guzzo, Marano, Marinoni, Mazure, Meneux,
  Merighi, Paltani, Pollo, Pozzetti, Radovich, Zucca, Bondi, Bongiorno,
  Busarello, {De La Torre}, Gregorini, Lamareille, Mathez, Merluzzi, Ripepi,
  Rizzo, \& Vergani}]{Ilbert2006}
Ilbert O. {et~al.}, 2006, Astronomy and Astrophysics, 457, 841

\bibitem[{Jarvis {et~al}\mbox{.}(2013)Jarvis, Bonfield, Bruce, Geach, McAlpine,
  McLure, Gonzalez-Solares, Irwin, Lewis, Yoldas, Andreon, Cross, Emerson,
  Dalton, Dunlop, Hodgkin, Le, Karouzos, Meisenheimer, Oliver, Rawlings,
  Simpson, Smail, Smith, Sullivan, Sutherland, White, \& Zwart}]{Jarvis2013}
Jarvis M.~J. {et~al.}, 2013, Monthly Notices of the Royal Astronomical Society,
  428, 1281

\bibitem[{Johnston {et~al}\mbox{.}(2015)Johnston, Vaccari, Jarvis, Smith,
  Giovannoli, H{\"{a}}u{\ss}ler, \& Prescott}]{Johnston2015}
Johnston R., Vaccari M., Jarvis M., Smith M., Giovannoli E., H{\"{a}}u{\ss}ler
  B., Prescott M., 2015, Monthly Notices of the Royal Astronomical Society,
  453, 2541

\bibitem[{Kauffmann {et~al}\mbox{.}(2013)Kauffmann, Li, Zhang, \&
  Weinmann}]{Kauffmann2013}
Kauffmann G., Li C., Zhang W., Weinmann S., 2013, Monthly Notices of the Royal
  Astronomical Society, 430, 1447

\bibitem[{Kawinwanichakij {et~al}\mbox{.}(2016)Kawinwanichakij, Quadri,
  Papovich, Kacprzak, Labb{\'{e}}, Spitler, Straatman, Tran, Allen, Behroozi,
  Cowley, Dekel, Glazebrook, Hartley, Kelson, Koo, Lee, Lu, Nanayakkara,
  Persson, Primack, Tilvi, Tomczak, \& van Dokkum}]{Kawinwanichakij2015}
Kawinwanichakij L. {et~al.}, 2016, The Astrophysical Journal, 817, 9

\bibitem[{Knobel {et~al}\mbox{.}(2012)Knobel, Lilly, Carollo, Contini, Kneib,
  {Le Fevre}, Mainieri, Renzini, Scodeggio, Zamorani, Bardelli, Bolzonella,
  Bongiorno, Caputi, Cucciati, de~la Torre, de~Ravel, Franzetti, Garilli,
  Iovino, Kampczyk, Kova{\v{c}}, Lamareille, {Le Borgne}, {Le Brun}, Maier,
  Mignoli, Pello, Peng, {Perez Montero}, Presotto, Silverman, Tanaka, Tasca,
  Tresse, Vergani, Zucca, Barnes, Bordoloi, Cappi, Cimatti, Coppa, Koekemoer,
  L{\'{o}}pez-Sanjuan, McCracken, Moresco, Nair, Pozzetti, \&
  Welikala}]{Knobel2012}
Knobel C. {et~al.}, 2012, The Astrophysical Journal, 755, 48

\bibitem[{Knobel {et~al}\mbox{.}(2015)Knobel, Lilly, Woo, \&
  Kova{\v{c}}}]{Knobel2014}
Knobel C., Lilly S.~J., Woo J., Kova{\v{c}} K., 2015, The Astrophysical
  Journal, 800, 24

\bibitem[{Landy \& Szalay(1993)}]{Landy1993}
Landy S.~D., Szalay A.~S., 1993, The Astrophysical Journal, 412, 64

\bibitem[{Lindsay {et~al}\mbox{.}(2014)Lindsay, Jarvis, \&
  McAlpine}]{Lindsay2014}
Lindsay S.~N., Jarvis M.~J., McAlpine K., 2014, Monthly Notices of the Royal
  Astronomical Society, 440, 2322

\bibitem[{Madau \& Dickinson(2014)}]{Madau2014}
Madau P., Dickinson M., 2014, Annual Review of Astronomy and Astrophysics, 52,
  415

\bibitem[{Madau {et~al}\mbox{.}(1996)Madau, Ferguson, Dickinson, Giavalisco,
  Steidel, \& Fruchter}]{Madau1996}
Madau P., Ferguson H.~C., Dickinson M.~E., Giavalisco M., Steidel C.~C.,
  Fruchter A., 1996, Monthly Notices of the Royal Astronomical Society, 283,
  1388

\bibitem[{Madau {et~al}\mbox{.}(1998)Madau, Pozzetti, \& Dickinson}]{Madau1998}
Madau P., Pozzetti L., Dickinson M., 1998, The Astrophysical Journal, 498, 106

\bibitem[{McAlpine {et~al}\mbox{.}(2012)McAlpine, Smith, Jarvis, Bonfield, \&
  Fleuren}]{McAlpine2012}
McAlpine K., Smith D. J.~B., Jarvis M.~J., Bonfield D.~G., Fleuren S., 2012,
  Monthly Notices of the Royal Astronomical Society, 423, 132

\bibitem[{McCracken {et~al}\mbox{.}(2015)McCracken, Wolk, Colombi, Kilbinger,
  Ilbert, Peirani, Coupon, Dunlop, Milvang-Jensen, Caputi, Aussel, Bethermin,
  \& {Le Fevre}}]{McCracken2015}
McCracken H.~J. {et~al.}, 2015, Monthly Notices of the Royal Astronomical
  Society, 449, 901

\bibitem[{McGee {et~al}\mbox{.}(2014)McGee, Bower, \& Balogh}]{McGee2014}
McGee S.~L., Bower R.~G., Balogh M.~L., 2014, Monthly Notices of the Royal
  Astronomical Society: Letters, 442, L105

\bibitem[{Navarro {et~al}\mbox{.}(1996)Navarro, Frenk, \& White}]{Navarro1996}
Navarro J.~F., Frenk C.~S., White S. D.~M., 1996, The Astrophysical Journal,
  462, 563

\bibitem[{{Oemler, Augustus}(1974)}]{Oemler1974}
{Oemler, Augustus} J., 1974, The Astrophysical Journal, 194, 1

\bibitem[{Oke \& Gunn(1983)}]{Oke1983}
Oke J.~B., Gunn J.~E., 1983, The Astrophysical Journal, 266, 713

\bibitem[{Pahwa \& Paranjape(2017)}]{Pahwa2017}
Pahwa I., Paranjape A., 2017, Monthly Notices of the Royal Astronomical
  Society, 470, 1298

\bibitem[{Paranjape {et~al}\mbox{.}(2015)Paranjape, Kova{\v{c}}, Hartley, \&
  Pahwa}]{Paranjape2015}
Paranjape A., Kova{\v{c}} K., Hartley W.~G., Pahwa I., 2015, Monthly Notices of
  the Royal Astronomical Society, 454, 3030

\bibitem[{Patel {et~al}\mbox{.}(2009)Patel, Holden, Kelson, Illingworth, \&
  Franx}]{Patel2009}
Patel S.~G., Holden B.~P., Kelson D.~D., Illingworth G.~D., Franx M., 2009, The
  Astrophysical Journal, 705, L67

\bibitem[{Peng {et~al}\mbox{.}(2010)Peng, Lilly, Kova{\v{c}}, Bolzonella,
  Pozzetti, Renzini, Zamorani, Ilbert, Knobel, Iovino, Maier, Cucciati, Tasca,
  Carollo, Silverman, Kampczyk, de~Ravel, Sanders, Scoville, Contini, Mainieri,
  Scodeggio, Kneib, {Le F{\`{e}}vre}, Bardelli, Bongiorno, Caputi, Coppa, de~la
  Torre, Franzetti, Garilli, Lamareille, {Le Borgne}, {Le Brun}, Mignoli,
  Montero, Pello, Ricciardelli, Tanaka, Tresse, Vergani, Welikala, Zucca,
  Oesch, Abbas, Barnes, Bordoloi, Bottini, Cappi, Cassata, Cimatti, Fumana,
  Hasinger, Koekemoer, Leauthaud, Maccagni, Marinoni, McCracken, Memeo, Meneux,
  Nair, Porciani, Presotto, \& Scaramella}]{Peng2010a}
Peng Y.-j. {et~al.}, 2010, The Astrophysical Journal, 721, 193

\bibitem[{Prescott {et~al}\mbox{.}(2011)Prescott, Baldry, James, Bamford,
  Bland-Hawthorn, Brough, Brown, Cameron, Conselice, Croom, Driver, Frenk,
  Gunawardhana, Hill, Hopkins, Jones, Kelvin, Kuijken, Liske, Loveday, Nichol,
  Norberg, Parkinson, Peacock, Phillipps, Pimbblet, Popescu, Robotham, Sharp,
  Sutherland, Taylor, Tuffs, van Kampen, \& Wijesinghe}]{Prescott2011}
Prescott M. {et~al.}, 2011, Monthly Notices of the Royal Astronomical Society,
  417, 1374

\bibitem[{Quadri {et~al}\mbox{.}(2012)Quadri, Williams, Franx, \&
  Hildebrandt}]{Quadri2012}
Quadri R.~F., Williams R.~J., Franx M., Hildebrandt H., 2012, The Astrophysical
  Journal, 744, 88

\bibitem[{Rawlings \& Jarvis(2004)}]{Rawlings2004}
Rawlings S., Jarvis M.~J., 2004, Monthly Notices of the Royal Astronomical
  Society, 355, L9

\bibitem[{Scoccimarro {et~al}\mbox{.}(2001)Scoccimarro, Sheth, Hui, \&
  Jain}]{Scoccimarro2001}
Scoccimarro R., Sheth R.~K., Hui L., Jain B., 2001, The Astrophysical Journal,
  546, 20

\bibitem[{Scoville {et~al}\mbox{.}(2007)Scoville, Abraham, Aussel, Barnes,
  Benson, Blain, Calzetti, Comastri, Capak, Carilli, Carlstrom, Carollo,
  Colbert, Daddi, Ellis, Elvis, Ewald, Fall, Franceschini, Giavalisco, Green,
  Griffiths, Guzzo, Hasinger, Impey, Kneib, Koda, Koekemoer, Lefevre, Lilly,
  Liu, McCracken, Massey, Mellier, Miyazaki, Mobasher, Mould, Norman,
  Refregier, Renzini, Rhodes, Rich, Sanders, Schiminovich, Schinnerer,
  Scodeggio, Sheth, Shopbell, Taniguchi, Tyson, Urry, {Van Waerbeke},
  Vettolani, White, \& Yan}]{Scoville2007}
Scoville N. {et~al.}, 2007, The Astrophysical Journal Supplement Series, 172,
  38

\bibitem[{Seljak(2009)}]{Seljak2009}
Seljak U., 2009, Physical Review Letters, 102, 021302

\bibitem[{Simon {et~al}\mbox{.}(2009)Simon, Hetterscheidt, Wolf, Meisenheimer,
  Hildebrandt, Schneider, Schirmer, \& Erben}]{Simon2008}
Simon P., Hetterscheidt M., Wolf C., Meisenheimer K., Hildebrandt H., Schneider
  P., Schirmer M., Erben T., 2009, Monthly Notices of the Royal Astronomical
  Society, 398, 807

\bibitem[{Skibba {et~al}\mbox{.}(2009)Skibba, Bamford, Nichol, Lintott,
  Andreescu, Edmondson, Murray, Raddick, Schawinski, Slosar, Szalay, Thomas, \&
  Vandenberg}]{Skibba2009}
Skibba R.~A. {et~al.}, 2009, Monthly Notices of the Royal Astronomical Society,
  399, 966

\bibitem[{Smethurst {et~al}\mbox{.}(2015)Smethurst, Lintott, Simmons,
  Schawinski, Marshall, Bamford, Fortson, Kaviraj, Masters, Melvin, Nichol,
  Skibba, \& Willett}]{Smethurst2015}
Smethurst R.~J. {et~al.}, 2015, Monthly Notices of the Royal Astronomical
  Society, 450, 435

\bibitem[{Szapudi \& Szalay(1998)}]{Szapudi1998}
Szapudi I., Szalay A.~S., 1998, The Astrophysical Journal, 494, L41

\bibitem[{Tinker {et~al}\mbox{.}(2017)Tinker, Hahn, Mao, Wetzel, \&
  Conroy}]{Tinker2017}
Tinker J.~L., Hahn C., Mao Y.-Y., Wetzel A.~R., Conroy C., 2017, eprint
  arXiv:1702.01121

\bibitem[{Tinker {et~al}\mbox{.}(2013)Tinker, Leauthaud, Bundy, George,
  Behroozi, Massey, Rhodes, \& Wechsler}]{Tinker2013a}
Tinker J.~L., Leauthaud A., Bundy K., George M.~R., Behroozi P., Massey R.,
  Rhodes J., Wechsler R.~H., 2013, The Astrophysical Journal, 778, 93

\bibitem[{Tonnesen \& Cen(2014)}]{Tonnesen2014}
Tonnesen S., Cen R., 2014, The Astrophysical Journal, 788, 133

\bibitem[{Viola {et~al}\mbox{.}(2015)Viola, Cacciato, Brouwer, Kuijken,
  Hoekstra, Norberg, Robotham, van Uitert, Alpaslan, Baldry, Choi, de~Jong,
  Driver, Erben, Grado, Graham, Heymans, Hildebrandt, Hopkins, Irisarri,
  Joachimi, Loveday, Miller, Nakajima, Schneider, Sif{\'{o}}n, \& {Verdoes
  Kleijn}}]{Viola2015}
Viola M. {et~al.}, 2015, Monthly Notices of the Royal Astronomical Society,
  452, 3529

\bibitem[{Weinberg \& Kamionkowski(2003)}]{Weinberg2003}
Weinberg N.~N., Kamionkowski M., 2003, Monthly Notices of the Royal
  Astronomical Society, 341, 251

\bibitem[{Weinmann {et~al}\mbox{.}(2006)Weinmann, van~den Bosch, Yang, \&
  Mo}]{Weinmann2006}
Weinmann S.~M., van~den Bosch F.~C., Yang X., Mo H.~J., 2006, Monthly Notices
  of the Royal Astronomical Society, 366, 2

\bibitem[{Welikala {et~al}\mbox{.}(2015)Welikala, Bethermin, Guery, Strandet,
  Aird, Aravena, Ashby, Bothwell, Beelen, Bleem, de~Breuck, Brodwin, Carlstrom,
  Chapman, Crawford, Dole, Dore, Everett, Flores-Cacho, Gonzalez,
  Gonzalez-Nuevo, Greve, Gullberg, Hezaveh, Holder, Holzapfel, Keisler,
  Lagache, Ma, Malkan, Marrone, Mocanu, Montier, Murphy, Nesvadba, Omont,
  Pointecouteau, Puget, Reichardt, Rotermund, Scott, Serra, Spilker, Stalder,
  Stark, Story, Vanderlinde, Vieira, \& Weiss}]{Welikala2015}
Welikala N. {et~al.}, 2015, Monthly Notices of the Royal Astronomical Society,
  455, 1629

\bibitem[{White {et~al}\mbox{.}(2015)White, Jarvis, Haussler, \&
  Maddox}]{White2015}
White S.~V., Jarvis M.~J., Haussler B., Maddox N., 2015, Monthly Notices of the
  Royal Astronomical Society, 448, 2665

\bibitem[{Woo {et~al}\mbox{.}(2013)Woo, Dekel, Faber, Noeske, Koo, Gerke,
  Cooper, Salim, Dutton, Newman, Weiner, Bundy, Willmer, Davis, \&
  Yan}]{Woo2012}
Woo J. {et~al.}, 2013, Monthly Notices of the Royal Astronomical Society, 428,
  3306

\bibitem[{Yang {et~al}\mbox{.}(2012)Yang, Mo, van~den Bosch, Zhang, \&
  Han}]{Yang2011}
Yang X., Mo H.~J., van~den Bosch F.~C., Zhang Y., Han J., 2012, The
  Astrophysical Journal, 752, 41

\bibitem[{Zu \& Mandelbaum(2015)}]{Zu2015}
Zu Y., Mandelbaum R., 2015, Monthly Notices of the Royal Astronomical Society,
  454, 1161

\bibitem[{Zu \& Mandelbaum(2016)}]{Zu2015a}
Zu Y., Mandelbaum R., 2016, Monthly Notices of the Royal Astronomical Society,
  457, 4360

\bibitem[{Zwart {et~al}\mbox{.}(2014)Zwart, Jarvis, Deane, Bonfield, Knowles,
  Madhanpall, Rahmani, \& Smith}]{Zwart2014}
Zwart J. T.~L., Jarvis M.~J., Deane R.~P., Bonfield D.~G., Knowles K.,
  Madhanpall N., Rahmani H., Smith D. J.~B., 2014, Monthly Notices of the Royal
  Astronomical Society, 439, 1459

\end{thebibliography}
%\bibdata{MasterReferences}

\bsp

\label{lastpage}

\end{document}